%% file: sequentialcorr.tex
\documentclass[11pt]{article}
\usepackage{times}
\usepackage{chicagor}
\usepackage{smallsec}
\usepackage{url}
\usepackage{epsf}
\input{defn}


\input{bghkmac}

\newcommand{\undefined}{\uparrow\!\!}

\renewcommand{\H}{H}

\renewcommand{\A}{{A}}

\renewcommand{\M}{M}

\newtheorem{FACT}[THEOREM]{Fact}
                            {\end{FACT}}

\renewcommand{\Sigma}{S}
\newcommand{\shortv}{\commentout}
\newcommand{\fullv}[1]{#1}
\newcommand{\EU}{{\mathrm{EU}}}

\DeclareMathSizes{12}{12}{10}{9}

\title{Generalized Solution Concepts in Games with Possibly Unaware Players}

\author{Leandro C. R\^ego\\
Statistics Department \\
Federal University of Pernambuco\\
Recife-PE, Brazil \\
e-mail: leandro@de.ufpe.br\\
\and Joseph Y. Halpern
\\ Computer Science Department \\
Cornell University\\
Ithaca-NY, U.S.A. \\
e-mail: halpern@cs.cornell.edu}

\commentout{
\titlenote{
This work was supported in part by NSF under grants CTC-0208535,
ITR-0325453, and IIS-0534064, by ONR under grants N00014-00-1-03-41
and N00014-01-10-511, and by the DoD Multidisciplinary University
Research Initiative (MURI) program administered by the ONR under
grant N00014-01-1-0795. The second author was also supported in part
by a scholarship from the Brazilian Government through the Conselho
Nacional de Desenvolvimento Cient\'ifico e Tecnol\'ogico (CNPq). } }

\begin{document}
\date{}
\maketitle
\thispagestyle{empty}

\begin{abstract}
Most work in game theory assumes that players are perfect
reasoners and have common knowledge of all significant aspects of
the game. In earlier work \cite{HR06}, we proposed a framework
for representing and analyzing games with possibly unaware players,
and suggested a generalization of Nash equilibrium appropriate for games
with unaware players that we called \emph{generalized Nash equilibrium}.
Here,
we use this framework to analyze other solution concepts%
\fullv{ that have
been considered in the game-theory literature}, with a focus on
sequential equilibrium.
We also provide some insight into the notion of generalized Nash equilibrium
by proving that it is closely related to the notion of
rationalizability when we restrict the analysis to games in normal
form and no unawareness is involved.
\end{abstract}

\section{INTRODUCTION}
\label{chap2:sec:intro}
Game theory has proved to be a useful tool in
the modeling and analysis of many phenomena involving interaction
between multiple agents. However, standard models used in game theory
implicitly assume that agents
are perfect reasoners and have common knowledge of all significant
aspects of the game. There are many situations where these
assumptions are not reasonable. In large games, agents may not be
aware of the other players in the game or all the moves a player can
make. Recently, we \cite{HR06} proposed a way of modeling such
games.
A key feature of this approach is the use of an {\em augmented
game}, which represents what players are aware of at each node of an
extensive form representation of a game.
Since the game is no longer assumed to be common knowledge, each
augmented game represents the game a player considers possible in
some situation and describes how he believes each other player's {\em
awareness level} changes over time, where intuitively the awareness
level of a player is the set of histories of the game that the
player is aware of.

In games with possibly unaware players, standard solution concepts
cannot be applied. For example, in a standard game a strategy
profile is a {\em Nash equilibrium} if each agent's strategy is a
best response to the other agents' strategies, so each agent $i$
would continue playing his strategy even if $i$ knew what strategies
the other agents were using. In the presence of unawareness this no
longer make sense, since the strategies used by other players may
involve moves $i$ is unaware of. We proposed a generalization of
Nash equilibrium consisting of a collection of strategies, one for
each pair $(i,\Gamma')$, where $\Gamma'$ is a game that agent $i$
considers to be the true game in some situation. Intuitively, the
strategy for a player $i$ at $\Gamma'$ is the strategy $i$ would
play in situations where $i$ believes that the true game is
$\Gamma'$. Roughly speaking, a generalized strategy profile
$\vec{\sigma}$, which includes a strategy $\sigma_{i,\Gamma'}$ for
each pair $(i,\Gamma')$, is a \emph{generalized Nash equilibrium} if
$\sigma_{i,\Gamma'}$ is a best response for player $i$ if the true
game is $\Gamma'$, given the strategies being used by the other
players in $\Gamma'$.
We showed that every game with awareness has a
generalized Nash equilibrium by associating a game with
awareness with a standard game (where agents are aware of all moves) and
proving
that there is a one-to-one correspondence between the generalized
Nash equilibria of the game with awareness and the Nash equilibria
of the standard game.

\begin{figure}[htb]
\centering \epsfxsize=9cm \epsffile{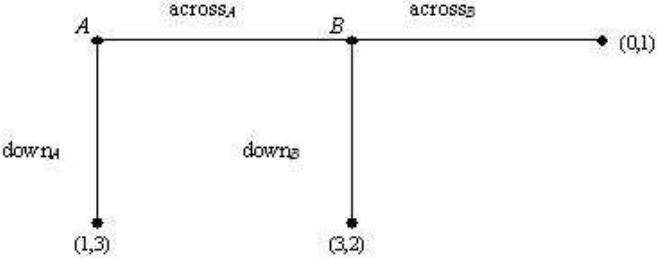} \caption{A simple
game.} \label{chap2:fig:game1}
\end{figure}

Some Nash equilibria seem unreasonable.
For example, consider the
game shown in Figure~\ref{chap2:fig:game1}.
\noindent
One Nash equilibrium of this game has $A$ playing down$_A$ and $B$
playing across$_B$.
In this equilibrium, $B$ gets a payoff of 3, and $A$ get a relatively
poor payoff of 1.  Intuitively, $A$ plays down because of $B$'s
``threat'' to play across$_B$.  But this threat does not appear to be so
credible.  If player $B$ is rational and ever gets
to move, he will not choose to move across$_B$ since it gives
him a lower payoff than playing down$_B$.
Moreover, if $B$ will play down$_B$ if he gets to move, then $A$ should
play across$_A$.
\commentout{
Our focus in this paper has been on refinements of generalized Nash
equilibrium in games with awareness.
It is worth reconsidering here the conceptual basis for these solution
concepts in the presence of awareness.%
\footnote{We thank Aviad Heifetz and an anonymous referee for raising
some of these issues.}
As we noted, equilibrium refinements are used in standard games to
eliminate some ``undesirable'' or ``unreasonable'' equilibria.
Arguably, unreasonable equilibria pose an even deeper problem with
unaware players.
For example,}

One standard interpretation of a Nash equilibrium is that a player
chooses his strategy at the beginning of the game, and then does not
change it because he has no motivation for doing so (since his payoff
is no higher when he changes strategies).  But this interpretation is
suspect in extensive-form games when a player makes a move that
takes the game off the equilibrium path.  It may seem unreasonable for a
player to then play
the move called for by his strategy (even if the strategy is part of a
Nash equilibrium), as in the case of player $B$ choosing across$_B$ in the example of Figure~\ref{chap2:fig:game1}.
In other words, a threat to blow up the world if I catch you
cheating in a game may be part of a Nash equilibrium, and does not cause
problems if in fact no one cheats, but it hardly seems credible if
someone does cheat.

One way to justify the existence of incredible
threats off the equilibrium path in a Nash equilibrium is to view the
player as choosing a computer program that will play the game for him,
and then leaving. Since the program is not changed once it is set in motion, threats about
moves that will be made at information sets off the equilibrium path become more credible.
However, in a game with awareness, a player cannot write a program to
play the whole game at the beginning of the game because, when his level of
awareness changes, he realizes that there are moves available to him
that he was not aware of at the beginning of the game.
He thus must write a new program that takes this into account.
But this means we cannot sidestep the problem of incredible threats by
appealing to the use of a pre-programmed computer to play a strategy.
Once we allow a player to change his program, threats that were made
credible because the program could not be rewritten become incredible
again.  Thus, the consideration of equilibrium refinements that block incredible threats becomes even
more pertinent with awareness.\footnote{We thank Aviad Heifetz and an anonymous referee for raising
some of these issues.}

\commentout{
One reason that Nash equilibrium fails to give a reasonable answer
in standard extensive games in general is because,
although each
player's strategy in a Nash equilibrium is a best response to the
other players' strategies, the move made at
an information set does not have to be a best response if that
information set is reached with probability 0. For
example, as we observed, in the game described in
Figure~\ref{chap2:fig:game1}, the profile where $A$ moves down and
$B$ moves across, which we denote $($down$_A$, across$_B)$, is a
Nash equilibrium. Nevertheless moving down is not a best response
for $B$ if $B$ is actually called upon to play.  The only reason
that this is a Nash equilibrium is that $B$ does not in fact
play,
since the information set where he plays is reached with probability 0,
given that $($down$_A$, across$_B)$
is the strategy profile.
}
There have been a number of variants of Nash equilibrium proposed in
the literature
to deal with this problem (and others), including
{\em perfect equilibrium} \cite{Selten75}, {\em proper
equilibrium} \cite{Myerson78},
{\em sequential equilibrium} \cite{KW82},
and {\em rationalizability} \cite{Ber84,Pearce84}, to name just a few.
Each of these solution concepts involves some notion of best response.
Our framework allows for straightforward generalizations of all
these solution concepts. As in our treatment of Nash equilibrium, if
$\Gamma_1 \ne \Gamma_2$, we treat player $i$ who considers the true
game to be $\Gamma_1$ to be a different agent from the version of
player $i$ who considers $\Gamma_2$ to be the true game. Each
version of player $i$ best responds (in the sense appropriate for
that solution concept) given his view of the game. In standard
games, it has been shown that, in each game, there is a strategy
profile satisfying that solution concept. Showing that an analogous
result holds in games with awareness can be nontrivial.
Instead of going through the process of generalizing every solution
concept, we focus here on {\em sequential equilibrium} since
(a) it is one of the best-known solution concepts for extensive games, (b)
the proof that a generalized sequential equilibrium exists suggests an interesting generalization of
sequential equilibrium for standard games, and (c) the techniques used to
prove its existence in games with awareness may generalize to
other solution concepts.

\fullv{Sequential equilibrium refines Nash
equilibrium (in the sense that every sequential equilibrium is a
Nash equilibrium) and does not allow solutions such as (down$_A$,
across$_B$). Intuitively, in a sequential equilibrium, every player
must make a best response at every information set (even if it is
reached with probability 0). In the game shown in
Figure~\ref{chap2:fig:game1}, the unique sequential equilibrium has $A$ choosing across$_A$ and $B$
choosing down$_B$.
We
propose a generalization of
sequential equilibrium to games with possibly unaware players,
and show that every game with awareness has a generalized sequential
equilibrium.
This turns out to be somewhat more subtle than the corresponding argument for generalized Nash equilibrium.
Our proof requires us to define a generalization of sequential equilibrium in standard games.
Roughly
speaking, this generalization relaxes the implicit assumption in
sequential equilibrium that every history in an information set is
actually considered possible by the player. We call this notion {\em
conditional sequential equilibrium}.
}
Other issues arise when considering sequential equilibrium in games with
awareness.  For example,
in a standard game, when a player reaches a history that is not on the
equilibrium path, he must believe that his opponent made a mistake.
However, in games with awareness,
a player may become aware of her own unawareness and, as a result,
switch strategies.
In the definition of sequential equilibrium in standard games, play off
the equilibrium path is dealt with by viewing it as the limit of ``small
mistakes'' (i.e., small deviations from the equilibrium strategy).
Given that there are alternative ways of dealing with mistakes in games
with awareness, perhaps other approaches for dealing with
off-equilibrium play might be more appropriate.
While other ways of dealing with mistakes may well prove interesting, we
would argue that our generalization of sequential equilibrium can be
motivated the same way as in standard games.
Roughly speaking, for us, how a player's awareness level changes over
time is not part of the equilibrium concept, but is given as part of the
description of the game.

We also provide some insight into the notion of generalized Nash
equilibrium by proving that, in a precise sense, it is closely related
to the notion of {\em rationalizability}
when we restrict the analysis to games in {\em
normal form} and no unawareness is involved (although the underlying
game is no longer common knowledge among the players). Roughly
speaking, a normal form game can be thought as a one-shot extensive
game where no player knows the move the others made before they make
their own move. Intuitively, in standard games, a strategy is
rationalizable for a player if it is a best response to some
reasonable beliefs he might have about the strategies being played
by other players, and a strategy is part of a Nash equilibrium if it
is a best response to the strategies actually played by the other
players. Since, in games with awareness, the game is not common
knowledge, a local strategy for player $i$ in $\Gamma^+$ is part of
a generalized Nash equilibrium if it is a best response to the
strategies played by the opponents of player $i$ in the games player
$i$ believes his opponents consider to be the actual one while
moving in $\Gamma^+$. Note that the line between rationalizability
and generalized Nash equilibrium is not sharp. In fact, we are
essentially able to prove that a strategy is rationalizable in a
standard game $\Gamma$ iff it is part of generalized Nash
equilibrium of an appropriate game with awareness whose underlying
game is $\Gamma$.

The rest of this paper is organized as follows. In
Section~\ref{sec:back}, we give the reader the necessary background
to understand this paper by reviewing our model for games with
awareness. In Section~\ref{chap2:sec:seq}, we review the definition
of sequential equilibrium for standard games and define its
generalization for games with awareness. In
Section~\ref{chap2:sec:stronseq}, we define the concept of
conditional sequential equilibrium for standard games, and prove
that there is a one-to-one correspondence between the generalized
sequential equilibria of a game with awareness and the conditional
sequential equilibria
of the standard game associated with it.
In Section~\ref{chap2:sec:grat}, we analyze the connection between
rationalizability and generalized Nash equilibrium.
We conclude in Section~\ref{chap2:sec:conc}.
Proofs of the theorems can be found in the Appendix.
\section{GAMES WITH AWARENESS}
\label{sec:back}

In this section, we introduce some notation and give some intuition
regarding games with awareness.
We encourage the reader to consult our earlier paper for details.

Games with awareness are modeled using \emph{augmented games}. Given
a standard extensive-form game described by a game tree $\Gamma$,
an augmented game $\Gamma^+$ \emph{based on} $\Gamma$
augments $\Gamma$ by describing each agent's \emph{awareness level}
at each node, where player $i$'s awareness level at a node $h$ is
essentially the set of \emph{runs} (complete histories) in $\Gamma$ that
$i$ is aware of at node $h$. A player's awareness level may change
over time, as the player becomes aware of more moves.

Formally, a \emph{(finite) extensive game} is a tuple $\Gamma=(N,\M,
\H,P,f_c,\{\I_i:i\in N\},\{u_i:i\in N\})$, where

\begin{itemize}

\item
\fullv{$N$ is a finite set consisting of the players of the game.}
\shortv{$N$ is a finite set of players.}

\item $M$ is a finite set whose elements are the
moves (or actions) available to players (and nature) during the
game.

\item $\H$ is a finite set of finite sequences of moves (elements of
$M$) that is closed under prefixes, so that if $h \in \H$ and $h'$
is a prefix of $h$, then $h' \in \H$. Intuitively, each member of
$\H$ is a \emph{history}.  We can identify the nodes in a game tree
with the histories in $\H$.  Each node $n$ is characterized by the
sequence of moves needed to reach $n$. A \emph{run} in $\H$ is a
terminal history, one that is not a strict prefix of any other
history in $\H$.  Let $Z$ denote the set of runs of $\H$. Let $M_h =
\{m \in M : h \cdot \<m\> \in \H\}$ (where we use $\cdot$ to denote
concatenation of sequences); $M_h$ is the set of moves that can be
made after history~$h$.

\item $P:(\H-Z)\rightarrow N\cup\{c\}$ is a function that assigns to
each nonterminal history $h$ a member of $N\cup\{c\}$. (We can think
of $c$ as representing nature.) If $P(h)=i$, then player $i$ moves
after history $h$; if $P(h)=c$, then nature moves after $h$. Let
$\H_i=\{h:P(h)=i\}$ be the set of all histories after which player
$i$ moves.

\item $f_c$ is a function that associates with every history for
which $P(h)=c$ a probability measure $f_c(\cdot \mid h)$ on $M_h$.
Intuitively, $f_c(\cdot\mid h)$ describes the probability of
nature's moves once history $h$ is reached.

\item $\I_i$ is a partition of
$\H_i$ with the property that if $h$ and $h'$ are in the same cell
of the partition then $M_h=M_{h'}$, i.e., the same set of moves is
available for every history in a cell of the partition. Intuitively,
if $h$ and $h'$ are in the same cell of $\I_i$, then $h$ and $h'$
are indistinguishable from $i$'s point of view; $i$ considers
history $h'$ possible if the actual history is $h$, and vice versa.
A cell $I\in\I_i$ is called an ($i$-)\emph{information set}.

\item $u_i:Z\rightarrow \mathrm{R}$ is a payoff function for player $i$,
assigning a real number ($i$'s payoff) to each run of the game.
\end{itemize}

An \emph{augmented game} is defined much like an extensive game. The
only essential difference is that at each nonterminal history we not
only determine the player moving but also her awareness level. There
are also extra moves of nature that intuitively capture players'
uncertainty regarding the awareness level of their opponents.
Formally, given an extensive game $\Gamma=(N, \M, \H,
P,f_c,\{\I_i:i\in N\},\{u_i:i\in N\})$, an {\em augmented game based
on $\Gamma$} is a tuple $\Gamma^+=(N^+, \M^+, \H^+,
P^+,f_c^+,\{\I_i^+:i\in N^+\},\{u_i^+:i\in N^+\},\{\A_i^+ :i\in
N^+\})$, where $(N^+, \M^+, \H^+, P^+,f_c^+,\{\I_i^+:i\in
N^+\},\{u_i^+:i\in N^+\})$ is a standard extensive game with perfect
recall%
\footnote{A game with perfect recall is one where, players remember all
the actions they have performed and all the information sets they have
passed through; see \cite{OR94} for the formal definition.}
and $\A_i^+:\H_i^+\rightarrow 2^\H$ describes $i$'s awareness
level at each history at which he moves. $\Gamma^+$ must satisfy
some consistency conditions. These conditions basically ensure that
\fullv{
\begin{itemize}
\item}
a player's awareness level depends only on the information she has\shortv{;}
\fullv{as captured by her information sets;}
\fullv{\item}
players do not forget histories that they were aware of;
and
\fullv{\item}
there is common knowledge of
(1) what the payoffs are
in the underlying game and (2) what the information sets are in the
underlying game.
\fullv{\end{itemize}}
The formal conditions are not needed in this paper, so we omit them here.

An augmented game describes either the modeler's view of the game or
the subjective view of the game of one of the players, and includes
both moves of the underlying game and moves of nature that change
awareness.
A game with awareness collects all these different views, and describes,
in each view, what view other players have.  Formally,
a \emph{game with awareness based on $\Gamma = (N, \M,\H,P,f_c,
\{\I_i:i\in N\},\{u_i: i\in N\})$} is a tuple $\Gamma^* = (\G,
\Gamma^m, \F)$, where
\begin{itemize}
\item $\G$ is a countable set of augmented games based on $\Gamma$, of
which one is $\Gamma^m$;
\item $\F$ maps an augmented game $\Gamma^+ \in\G$ and a history $h$ in
$\Gamma^+$ such that $P^+(h)=i$ to a pair $(\Gamma^h,I)$, where
$\Gamma^h\in \G$ and $I$ is an $i$-information set in game
$\Gamma^h$.
\end{itemize}
Intuitively, $\Gamma^m$ is the game from the point of view of an
omniscient modeler. If player $i$ moves at $h$ in game $\Gamma^+ \in
\G$ and $\F(\Gamma^+,h) = (\Gamma^h,I)$, then $\Gamma^h$ is the game
that $i$ believes to be the true game when the history is $h$, and
$I$ consists of the set of histories  in $\Gamma^h$ that $i$ currently
considers possible.

The augmented game $\Gamma^m$ and the mapping $\F$ must satisfy a
number of consistency conditions. The conditions on the modeler's
game ensures that the modeler is aware of all the players and moves
of the underlying game, and that he understands how nature's moves
work in the underlying game $\Gamma$. The game $\Gamma^m$ can be
thought of as a description of ``reality''; it describes the effect
of moves in the underlying game and how players' awareness levels
change. The other games in $\G$ describe a player's subjective view
of the situation.
There are also ten constraints on the mapping $\F$ that capture
desirable properties of awareness.
Rather than describing all ten constraints here, we briefly describe
a few of them, to give some idea of
the intuition behind these constraints. Suppose that
$\F(\Gamma^+,h) = (\Gamma^h,I)$ and $\A_i^+(h)=a$,\footnote{As in our earlier paper, we use the convention
that the components
of a (standard or augmented) game $\Gamma^+$ are labeled with the same superscript $+$, so that
we have $\M^+$, $\H^+$, $\A_i^+$, and so on.} then
the following conditions hold.
\begin{itemize}
\item[C1.]$\{\overline{h}: h \in \H^h\} = a$, where $\overline{h}$
is the subsequence of $h$ consisting of all the moves
in $h$ that are also in the set of moves $M$ of the underlying game $\Gamma$.

\item[C2.] If $h' \in \H^h$ and $P^h(h') = j$, then $\A_j^h(h')
\subseteq a$ and $M_{\overline{h}'}\inter \{m: \overline{h}'\cdot\<
m\>\in a\}= M^h_{h'}$.

\item[C5.] If $h' \in \H^+$, $P^+(h') = i$, $\A_i^+(h')=a$,
then if $h$ and $h'$ are in the same information set of $\Gamma^+$,
then $\F(\Gamma^+,h') = (\Gamma^{h},I)$, while if $h$ is a prefix or
a suffix of $h'$, then $\F(\Gamma^+,h') = (\Gamma^{h},I')$ for some
$i$-information set $I'$.

\item[C8.] For all histories $h'\in I$, there exists a prefix $h'_1$ of
$h'$ such that $P^h(h'_1)=i$ and $\F(\Gamma^h,h'_1)=(\Gamma',I')$
iff there exists a prefix $h_1$ of $h$ such that $P^+(h_1)=i$ and
$\F(\Gamma^+,h_1)=(\Gamma',I')$. Moreover, $h'_1\cdot \<m\>$ is a
prefix of $h'$ iff $h_1\cdot \<m\>$ is a prefix of $h$.

\item[C9.] There exists
a history $h'\in I$ such that for every prefix $h''\cdot \<m\>$ of
$h'$, if $P^h(h'')=j \in N^h$ and $\F(\Gamma^h,h'')=(\Gamma',I')$,
then for all $h_1\in I'$, $h_1\cdot \<m\>\in \H'$.
\end{itemize}

Suppose that $\F(\Gamma^+,h) = (\Gamma^h,I)$. Player $i$ moving at
history $h$ in $\Gamma^+$ thinks the actual game is $\Gamma^h$.
Moreover, $i$ thinks he is in the information set of $I$ of
$\Gamma^h$. C1 guarantees that the set of histories of the
underlying game player $i$ is aware of is exactly the set of
histories of the underlying game that appear in $\Gamma^h$. C2
states that no player in $\Gamma^h$ can be aware of histories not in
$a$. The second part of C2 implies that the set of moves available
for player $j$ at $h'$ is just the set of moves that player $i$ is
aware of that are available for $j$ at $\overline{h}'$ in the
underlying game. C5 says that player $i$'s subjective view of the
game changes only if $i$ becomes aware of more moves and is the same
at histories in $\H^+$ that $i$ cannot distinguish.

C8 is a consequence of the perfect recall assumption. C8 says that
if, at history $h$, $i$ considers $h'$ possible, then for every
prefix $h_1'$ of $h'$ there is a corresponding prefix of $h$ where
$i$ considers himself to be playing the same game, and similarly,
for every prefix of $h$ there is a prefix of $h'$ where $i$
considers himself to be playing the same game.  Moreover, $i$ makes
the same move at these prefixes. The intuition behind condition C9
is that player $i$ knows that player $j$ only make moves that $j$ is
aware of. Therefore, player $i$ must consider at least one history
$h'$ where he believes that every player $j$ made a move that $j$
was aware of. It follows from conditions on augmented games, C1, C2,
and C9 that there is a run going through $I$ where every player $j$
makes a move that player $i$ believes that $j$ is aware of.
It may seem that by making $\F$ a function we cannot capture a
player's uncertainty about the game being played or uncertainty about
opponents' unawareness about histories. However, we can
capture such uncertainty by folding it into nature's initial move in the
game the player consider possible while moving. It should be clear that
this gives a general approach to capturing such uncertainties.

We identify a standard extensive game $\Gamma$ with the game
$(\{\Gamma^m\},\Gamma^m,\F)$, where (abusing notation slightly)
$\Gamma^m = (\Gamma,\{\A_i: i \in N\})$ and, for all histories $h$
in an $i$-information set $I$ in $\Gamma$, $\A_i(h) = \H$ and
$\F(\Gamma^m,h) = (\Gamma^m,I)$. Thus, all players are aware of all
the runs in $\Gamma$, and agree with each other and the modeler that
the game is $\Gamma$. This is the \emph{canonical representation of
$\Gamma$} as a game with awareness.

In \cite{HR06}, we discussed generalizations of games with awareness to
include situations where players may be aware of their own unawareness
and, more generally, games where players may  not have common knowledge of
the underlying game is; for example, players may disagree
about what the
payoffs or the information sets are.  With these models, we can capture
a situation where, for example, player $i$ may think that another player $j$
cannot make a certain a certain move, when in fact $j$ can make such a
move.
For ease of exposition, we do not discuss these generalizations further
here.  However, it is not hard to show that the results of this paper can be
extended to them in a straightforward way.

Feinberg~\citeyear{Feinberg04,Feinberg05} also studied
games with awareness. Feinberg~\citeyear{Feinberg05} gives a
definition of extended Nash equilibrium in normal-form games.
Feinberg~\citeyear{Feinberg04} deals with extensive-form games and
defines solution concepts only indirectly, via a syntactic epistemic
characterization. His approach lacks a more direct semantic
framework, which our model provides. Li \citeyear{LI06b} has also
provided a model of unawareness in
extensive games, based on her earlier work on modeling
unawareness \cite{LI06,LI06a}.
See \cite{HR06} for some further discussion of the relation between these
approaches and ours.

\section{GENERALIZED SEQUENTIAL EQUILIBRIUM}
\label{chap2:sec:seq}

To explain generalized sequential equilibrium, we first review
the notion of sequential equilibrium for standard games.
\fullv{\subsection{SEQUENTIAL EQUILIBRIUM FOR STANDARD GAMES}}
Sequential equilibrium is defined with respect to an {\em assessment}, a pair
$(\vec{\sigma},\mu)$ where $\vec{\sigma}$ is a strategy profile
consisting of \emph{behavioral strategies}
and $\mu$ is a {\em belief system}, i.e., a function that determines for
every information set $I$ a probability $\mu_I$ over the histories
in $I$. Intuitively, if $I$ is an information set for player $i$,
$\mu_I$ is $i$'s subjective assessment of the relative likelihood of
the histories in $I$. Roughly speaking, an assessment is a
sequential equilibrium if for all players $i$, at every
$i$-information set, (a) $i$ chooses a best response given the
beliefs he has about the histories in that information set and the
strategies of other players, and (b) $i$'s beliefs are consistent
with the strategy profile being played, in the sense that they are
calculated by conditioning the probability distribution induced by
the strategy profile over the histories on the information set.

Note that $\mu_I$ is defined even if $I$ is reached with probability
0. Defining consistency at an information set that is reached with
probability 0 is somewhat subtle. In that case, intuitively, once
information set $I$ is reached player $i$ moving at $I$ must believe
the game has been played according to an alternative strategy
profile. In a sequential equilibrium, that alternative strategy
profile consists of a small perturbation of the original assessment
where every move is chosen with positive probability.

Given a strategy profile $\vec{\sigma}$, let $\Pr_{\vec{\sigma}}$ be
the probability distribution induced by $\vec{\sigma}$ over the
possible histories of the game. Intuitively, $\Pr_{\vec{\sigma}}(h)$
is the product of the probability of each of the moves in $h$. For
simplicity we assume $f_c>0$, so that if $\vec{\sigma}$ is such that
every player chooses all of his moves with positive probability,
then for every history $h$, $\Pr_{\vec{\sigma}}(h)>0$.%
\footnote{See
\cite{Myerson} for a definition of sequential equilibrium in
the case nature chooses some of its move with probability 0.} For
any history $h$ of the game, define $\Pr_{\vec{\sigma}}(\cdot\mid h)$
to be the conditional probability distribution induced by
$\vec{\sigma}$ over the possible histories of the game given that
the current history is $h$. Intuitively, $\Pr_{\vec{\sigma}}(h'\mid
h)$ is 0 if $h$ is not a prefix of $h'$, is 1 if $h=h'$, and is the
product of the probability of each of the moves in the path from $h$
to $h'$ if $h$ is a prefix of $h'$. Formally, an assessment
$(\vec{\sigma},\mu)$ is a sequential equilibrium if it satisfies the
following properties:
\begin{itemize}
\item {\em Sequential rationality.} For every information set $I$
and player $i$ and every behavioral strategy $\sigma$ for player $i$,
$$\EU_i((\vec{\sigma},\mu)\mid I)\geq
\EU_i(((\vec{\sigma}_{-i},\sigma),\mu)\mid I),$$
where $\EU_i((\vec{\sigma},\mu)\mid I)=\sum_{h\in I}\sum_{z\in
Z}\mu_I(h)\Pr_{\vec{\sigma}}(z\mid h)u_i(z)$.

\item {\em Consistency between belief system and strategy profile.}
If $\vec{\sigma}$ consists of \emph{completely mixed} (behavior)
strategies, that is, ones that assign positive probability to every
action at every information set, then for every information set $I$ and
history $h$ in $I$,
$$\mu_I(h)=\frac{\Pr_{\vec{\sigma}}(h)}{\sum_{h'\in
I}\Pr_{\vec{\sigma}}(h')}.$$

Otherwise, there exists a sequence $(\vec{\sigma}^n,\mu^n)$,
$n=1,2,3,\ldots$, of assessments such that $\vec{\sigma}^n$ consists
of completely mixed strategies,
$(\vec{\sigma}^n,\mu^n)$ is consistent in the above sense, and
$\lim_{n \tendsto \infty}(\vec{\sigma}^n,\mu^n)=(\vec{\sigma},\mu)$.
\end{itemize}

\fullv{
Sequential equilibrium is not a reasonable solution concept for
games with awareness for the same reason that Nash equilibrium is
not a reasonable solution concept for games with awareness; it
requires that a player be aware of the set of possible strategies
available to other players and to him.}
In order to define a
generalized notion of sequential equilibrium for games with
awareness, we first need to define a generalized notion of
assessment for games with awareness.
We first need a generalized notion of strategy, which we defined in our
earlier paper.

Intuitively, a
strategy describes what $i$ will do in every possible situation that
can arise. This no longer makes sense in games with awareness, since
a player no longer understands in advance all the possible
situations that can arise. For example, player $i$ cannot plan in
advance for what will happen if he becomes aware of something he is
initially unaware of. We solved this problem in our earlier paper as
follows. Let $\G_i=\{\Gamma'\in\G:\, \mbox{for some } \Gamma^+\in\G
\mbox{ and }h\mbox{ in }\Gamma^+, \, P^+(h)=i\mbox{ and
}\F(\Gamma^+,h)=(\Gamma',\cdot)\}$. Intuitively, $\G_i$ consists of
the games that $i$ views as the real game in some history. Rather
than considering a single strategy in a game $\Gamma^* = (\G,
\Gamma^m,\F)$ with awareness, we considered a collection
$\{\sigma_{i,\Gamma'}: \Gamma'\in \G_i \}$ of \emph{local
strategies}. Intuitively, a local strategy $\sigma_{i,\Gamma'}$ for
game $\Gamma'$ is the strategy that $i$ would use if $i$ were called
upon to play and $i$ thought that the true game was $\Gamma'$. Thus,
the domain of $\sigma_{i,\Gamma'}$ consists of pairs $(\Gamma^+,h)$
such that $\Gamma^+ \in \G$, $h$ is a history in $\Gamma^+$, $P^+(h)
= i$, and $\F(\Gamma^+,h) = (\Gamma',I)$.
Let $(\Gamma^h,I)^* = \{(\Gamma',h): \F(\Gamma',h)=(\Gamma^h,I)\}$; we
call $(\Gamma^h,I)^*$ a {\em generalized information set}.

\begin{definition} Given a game $\Gamma^* = (\G,\Gamma^m,\F)$ with awareness,
a \emph{local strategy} $\sigma_{i,\Gamma'}$ for agent $i$ is a
function mapping pairs $(\Gamma^+,h)$ such that $h$ is a history
where $i$ moves in $\Gamma^+$ and $\F(\Gamma^+,h) = (\Gamma',I)$ to
a probability distribution over $M'_{h'}$, the moves available at a
history $h' \in I$, such that $\sigma_{i,\Gamma'}(\Gamma_1,h_1) =
\sigma_{i,\Gamma'}(\Gamma_2,h_2)$ if $(\Gamma_1,h_1)$ and
$(\Gamma_2,h_2)$ are in the same generalized information set. A {\em generalized strategy profile} of $\Gamma^* =
(\G,\Gamma^m,\F)$ is a set of local strategies $\vec{\sigma} =
\{\sigma_{i,\Gamma'}:i\in N,\Gamma'\in\G_i\}$.
\fullv{
\end{definition}

The belief system, the second component of the assessment, is a
function from information sets $I$ to probability distribution over
the histories in $I$. Intuitively, it captures how likely each of
the histories in $I$ is for the player moving at $I$. For standard
games this distribution can be arbitrary, since the player considers
every history in the information set possible. This is no longer
true in games with awareness. It is possible that a player is
playing game $\Gamma_1$ but believes he is playing a different game
$\Gamma_2$. Furthermore, in an augmented game, there may be some
histories in an $i$-information set that include moves of which $i$
is not aware; player $i$ cannot consider these histories possible.
To deal with these problems, we define $\mu$ to be a {\em
generalized belief system} if it is a function from generalized
information sets to a probability distribution over the set of
histories in the generalized information set that the player
considers possible.

\begin{definition}
}
A {\em generalized belief system} $\mu$ is a function that
associates each generalized information set $(\Gamma', I)^*$ with a
probability distribution $\mu_{\Gamma',I}$ over the set
$\{(\Gamma',h):h\in I\}$.
A {\em generalized assessment} is a pair
$(\vec{\sigma},\mu)$, where $\vec{\sigma}$ is a generalized strategy
profile and $\mu$ is a generalized belief system.
\end{definition}

\commentout{
We want to define a notion of generalized sequential equilibrium
that captures the intuition that for every player $i$ and every
$i$-information set $I$, if $i$ thinks he is actually playing in the
information set $I$ of game $\Gamma'$, then his local strategy
$\sigma_{i,\Gamma'}$ is a best response to the local strategies of
other players in~$\Gamma'$ and his beliefs about the likelihood of
the histories in $I$, even if $I$ is reached with probability 0.

Let $\EU_{i,\Gamma'}((\vec{\sigma},\mu)\mid I)$ be the conditional
expected payoff for $i$ in the game $\Gamma'$ given that strategy
profile $\vec{\sigma}$ is used, information set $I$ has been
reached, and player $i$'s beliefs about the histories in $I$ are
described by $\mu_{\Gamma',I}$.
As in the case of generalized Nash
equilibrium, the only strategies in $\vec{\sigma}$ that are needed
to compute $\EU_{i,\Gamma'}((\vec{\sigma},\mu)\mid I)$ are the
strategies actually used in $\Gamma'$; indeed, all that is needed is
the restriction of these strategies to information sets that arise
in $\Gamma'$. We also do not need the whole generalized belief
system; only $\mu_{\Gamma',I}$ is needed.
A \emph{generalized sequential equilibrium} of $\Gamma^* =
(\G,\Gamma^m,\F)$ is a generalized assessment $(\vec{\sigma}^*,\mu^*)$
such that for every generalized information set $(\Gamma',I)^*$, the
local strategy $\sigma^*_{i,\Gamma'}$ is a best response to
$\vec{\sigma}^*_{-(i,\Gamma')}$ given $i$'s beliefs about the histories in
$I$, where $\vec{\sigma}^*_{-(i,\Gamma')}$ is the set of all local
strategies in $\vec{\sigma}^*$ other than $\sigma_{i,\Gamma'}^*$, and
$\mu^*$ is consistent with $\vec{\sigma}^*$. More formally, $(\vec{\sigma}^*,\mu^*)$
must satisfy the following two conditions
\begin{itemize}
\item {\em Generalized sequential rationality.} For every player $i$, generalized $i$-information set $(\Gamma',I)^*$, and local
strategy $\sigma$ for $i$ in $\Gamma'$,
$$\EU_{i,\Gamma'}((\vec{\sigma}^*,\mu^*)\mid I)\geq
\EU_{i,\Gamma'}(((\vec{\sigma}^*_{-(i,\Gamma')},\sigma),\mu^*)\mid
I),$$ where $\EU_{i,\Gamma'}((\vec{\sigma}^*,\mu^*)\mid I)=\sum_{h\in
I}\sum_{z\in Z'}\mu^*_{\Gamma',I}(h)\Pr_{\vec{\sigma}^*}(z\mid
h)u_i'(z)$.

\item {\em Consistency between generalized belief system and generalized strategy profile.}
If, for every generalized information set $(\Gamma',I)^*$,
$\sum_{h\in I}\Pr_{\vec{\sigma}^*}(h)>0$, then for all $h\in I$
$$\mu^*_{\Gamma',I}(h)=\frac{\Pr_{\vec{\sigma}^*}(h)}{\sum_{h'\in
I}\Pr_{\vec{\sigma}^*}(h')}.$$

Otherwise, there exists a sequence of generalized assessments
$(\vec{\sigma}^i,\mu^i)$ such that every player chooses all of his
moves with positive probability, $\mu^i$ is consistent with
$\vec{\sigma}^i$ and
$\lim_{i}(\vec{\sigma}^i,\mu^i)=(\vec{\sigma},\mu)$.
\end{itemize}
}

We can now define what it means for a generalized assessment
$(\vec{\sigma}^*,\mu^*)$ to be a
\emph{generalized sequential equilibrium} of a game with awareness.  The
definition is essentially identical to that of $(\vec{\sigma},\mu)$
being a sequential equilibrium;
the use of $\EU_i$ in the definition
of sequential rationality is replaced by $\EU_{i,\Gamma'}$, where
$\EU_{i,\Gamma'}((\vec{\sigma}^*,\mu^*)\mid I)$ is the conditional
expected payoff for $i$ in the game $\Gamma'$, given that strategy
profile $\vec{\sigma}^*$ is used, information set $I$ has been
reached, and player $i$'s beliefs about the histories in $I$ are
described by $\mu^*_{\Gamma',I}$.   We leave the straightforward
modifications of the definition to the reader.
It is easy to see that $(\vec{\sigma},\mu)$ is a sequential
equilibrium of a standard game $\Gamma$ iff $(\vec{\sigma},\mu)$ is a
(generalized) sequential equilibrium of the canonical representation
of $\Gamma$ as a game with awareness. Thus, our definition of
generalized sequential equilibrium generalizes the standard
definition.

\begin{figure}[htb]
\centering \epsfxsize=6cm \epsffile{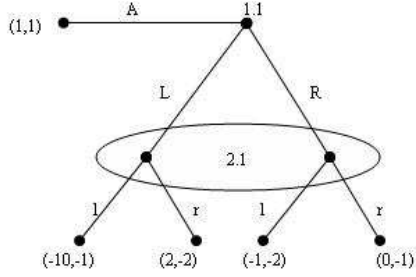} \caption{The
modeler's game $\Gamma^m$.} \label{chap2:fig:sequen_gammam}
\end{figure}

\begin{figure}[htb]
\centering \epsfxsize=6cm \epsffile{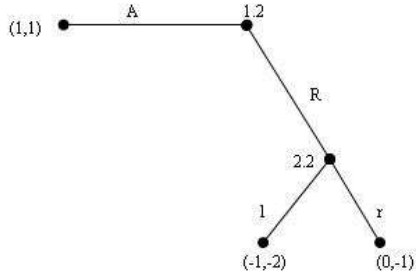}
\caption{Player 2's view of the game from point of view of player
1.} \label{chap2:fig:sequen_gamma22}
\end{figure}

To better understand the concept of generalized sequential
equilibrium concept, consider the game shown in
Figure~\ref{chap2:fig:sequen_gammam}.
Suppose that both players 1 and 2 are aware of all runs of the game, but
player 1 (falsely) believes that player 2 is aware only of the runs
not involving $L$ and believes that player 1 is aware of these runs
as well. Also suppose that player 2 is aware of all of this; that is,
player 2's view of the game is the same as the modeler's view of the
game $\Gamma^m$ shown in Figure~\ref{chap2:fig:sequen_gammam}.
While moving at node 1.1, player 1 considers the true game to be
identical to the modeler's game except that from player 1's point of
view, while moving at 2.1, player 2 believes the true game is
$\Gamma^{2.2}$, shown in Figure~\ref{chap2:fig:sequen_gamma22}.

This game has a unique generalized sequential equilibrium where
player 2 chooses $r$ and player 1 chooses $A$ in $\Gamma^{2.2}$.
Believing that player 2 will move $r$, player 1 best responds by
choosing $L$ at node 1.1. Since player 2 knows all this at node 2.1
in $\Gamma^m$, she chooses $l$ at this node. Thus, if players follow
their equilibrium strategies, the payoff vector is $(-10,-1)$. In
this situation, player 2 would be better off is she could let player
1 know that she is aware of move $L$, since then player 1 would play
$A$ and both players would receive 1. On the other hand, if we
slightly modify the game by making $u_2(\<L,l\>)=3$, then player 2
would benefit from the fact that 1 believes that she is unaware of
move $L$.
\fullv{\subsection{EXISTENCE OF GENERALIZED EQUILIBRIA}}
We now want to show that every game with awareness $\Gamma^*$
has at least one generalized sequential equilibrium. To prove that a
game with awareness
$\Gamma^*$ has a generalized Nash equilibrium, we constructed a standard
game $\Gamma^{\nu}$ with perfect recall and showed that there
exists a one-to-one correspondence between the set of generalized
Nash equilibrium of $\Gamma^*$ and the set of Nash equilibrium of
$\Gamma^{\nu}$.
Intuitively, $\Gamma^\nu$ is constructed by essentially ``gluing
together'' all the games $\Gamma' \in \G$, except that only
histories in $\Gamma'$ that can actually be played according to the
players' awareness level are considered.
More formally, given a game $\Gamma^* = (\G, \Gamma^m, \F)$ with awareness, let
$\nu$ be a probability on $\G$ that assigns each game in $\G$
positive probability.  (Here is where we use the fact that $\G$
is countable.)
For each $\Gamma'\in \G$, let $\lfloor
H^{\Gamma'}\rfloor=\{h\in H^{\Gamma'}:$ for every prefix $h_1\cdot
\<m\>$ of $h$, if $P'(h_1)=i \in N$ and
$\F(\Gamma',h_1)=(\Gamma'',I)$, then for all $h_2\in I$,
$h_2\cdot\<m\>\in \H''\}$.
The histories in $\lfloor
H^{\Gamma'}\rfloor$ are the ones that can actually be played according to the
players' awareness levels.

Let $\Gamma^\nu$ be the standard game such that
\begin{itemize}
\item $N^\nu = \{(i,\Gamma'):\Gamma'\in\G_i\}$;

\item $\M^\nu = \G\union_{\Gamma'\in\G}\lfloor M^{\Gamma'}\rfloor$, where
$\lfloor M^{\Gamma'}\rfloor$ is the set of moves that occur in
$\lfloor H^{\Gamma'}\rfloor$;

\item $\H^\nu = \<\, \>\union\{\<\Gamma'\>\cdot h:\Gamma'\in\G, h\in
\lfloor H^{\Gamma'}\rfloor\}$;

\item $P^\nu(\<\, \>)=c$, and $$P^\nu(\<\Gamma^h\>\cdot h') =
\left\{ \begin{array}{ll} (i,\Gamma^{h'}) &\mbox{if $P^h(h') = i \in
N$ and}\\
\ & \F(\Gamma^h,h')=(\Gamma^{h'}, \cdot),\\
c &\mbox{if $P^h(h') = c$;}\end{array} \right.$$

\item $f_c^\nu(\Gamma'|\<\, \>)= \nu(\Gamma')$ and
$f_c^\nu(\cdot|\<\Gamma^h\>\cdot h') = f_c^h(\cdot|h')$ if $P^h(h')
= c$;

\item $\I^\nu_{i,\Gamma'}$ is a partition of $\H^{\nu}_{i,\Gamma'}$ where two histories $\<\Gamma^1\>\cdot h^1$ and $\<\Gamma^1\>\cdot h^1$
are in the same information set $\<\Gamma',I\>^*$ iff
$(\Gamma^{1},h^1)$ and $(\Gamma^{2},h^2)$ are in the same
generalized information set $(\Gamma',I)^*$;

\item $u_{i,\Gamma'}^\nu(\<\Gamma^h\>\cdot z)= \left\{
\begin{array}{ll}
u_i^h(z) &\mbox{if $\Gamma^h = \Gamma',$}\\
0 &\mbox{if $\Gamma^h \ne \Gamma'.$}\end{array} \right.$
\end{itemize}

Unfortunately, while it is the case that there is a 1-1 correspondence between
the Nash equilibria of $\Gamma^{\nu}$ and the generalized Nash equilibria of $\Gamma^*$,
\shortv{as we show in the full paper},
this correspondence breaks down for sequential equilibria.
\shortv{Thus, we need to take a different approach to showing that
a generalized sequential equilibrium always exists.}
\fullv{
To see why consider the modified version of prisoner's dilemma $\Gamma^p$,
described in Figure~\ref{chap2:fig:game2}.

\begin{figure}[htb]
\centering \epsfxsize=9cm \epsffile{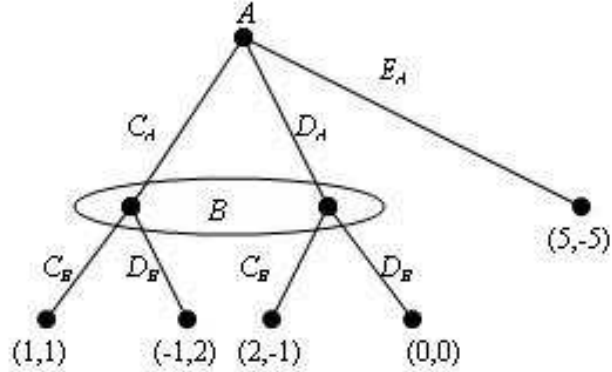}
\caption{$\Gamma^p$: modified prisoner's dilemma.} \label{chap2:fig:game2}
\end{figure}

Besides being able
to cooperate ($C_A$) or defect ($D_A$), player $A$ who moves first
has also the option of escaping ($E_A$). If player $A$ escapes, then
the game is over; if player $A$
cooperates or defects, then player $B$ may also cooperate ($C_B$) or
defect ($D_B$).
Suppose further that in the modeler's game,
\begin{itemize}
\item both $A$ and $B$ are aware of all histories of
$\Gamma^p$;

\item with probability $p$, $A$ believes
that $B$ is unaware of the extra move $E_A$, and with probability
$1-p$, $A$ believes $B$ is aware of all histories;

\item if $A$ believes $B$ is unaware of $E_A$, then $A$ believes
that $B$ believes that it is common knowledge that the game being
played contains all histories but $E_A$;

\item if $A$ believes $B$ is aware of $E_A$, then $A$ believes
that $B$ believes that there is common knowledge that the game being
played is $\Gamma^p$;

\item $B$ believes that it is common knowledge that the game being played is $\Gamma^p$.
\end{itemize}

We need four augmented games to model this situation:
\begin{itemize}
\item $\Gamma^m$ is the game from the modeler's point of view;

\item $\Gamma^A$ is the game from $A$'s point of view when she is called
to move in the modeler's game;

\item $\Gamma^{B.1}$ is the game from $B$'s point of view when he is
called to move in game $\Gamma^A$ after nature chooses he is unaware
of $E_A$ and is also the game from $A$'s point of view when she is
called to move in $\Gamma^{B.1}$; and

\item $\Gamma^{B.2}$ is the game from $B$'s point of view when he is
called to move in game $\Gamma^A$ after nature chooses aware$_B$;
$\Gamma^{B.2}$ is also the game from $B$'s point of view when he is
called to move at $\Gamma^m$ and the game from $A$'s point of view
when she is called to move in $\Gamma^{B.2}$.
\end{itemize}

Although $\Gamma^m$ and $\Gamma^{B.2}$ have the same game tree as
$\Gamma^p$, they are different augmented games, since the $\F$
function is defined differently at histories in these games.
For example, $\F(\Gamma^m,\<\
\>)=(\Gamma^A,\{$unaware$_B$,aware$_B\})\ne (\Gamma^{B.2},\<\
\>)=\F(\Gamma^{B.2},\<\ \>)$. For this reason, we use different labels for the nodes
of theses games. Let $A.3$ and $B.2$ (resp.,
$A.2$ and $B.2$) be the labels of the nodes in game $\Gamma^m$
(resp., $\Gamma^{B.2}$) corresponding to $A$ and $B$ in $\Gamma^p$,
respectively.  $\Gamma^A$ and $\Gamma^{B.1}$ are
shown in Figures~\ref{chap2:fig:gammaA} and \ref{chap2:fig:gammaB1},
respectively.%
\footnote{We abuse notation and use the same label for nodes in
different augmented
games that are in the same generalized information set. For example, $A.3$ is a label at both $\Gamma^m$ and $\Gamma^A$.}

\begin{itemize}
\item In the modeler's game $\Gamma^m$,
$A$ believes she is
playing game $\Gamma^{A}$, and
$B$ believes he is playing game $\Gamma^{B.2}$.

\item In game $\Gamma^{A}$, nature chooses move unaware$_B$ with
probability $p$ and aware$_B$ with probability $1-p$. Then $A$ moves
and believes she is playing $\Gamma^{A}$. At node $B.1$, $B$
believes he is playing $\Gamma^{B.1}$, and at node $B.2$, $B$
believes he is playing $\Gamma^{B.2}$.

\item In game $\Gamma^{B.1}$, $A$
and $B$ both believe that
the game is $\Gamma^{B.1}$.

\item In game $\Gamma^{B.2}$,
$A$ and $B$ both believe that the game is $\Gamma^{B.2}$.
 \end{itemize}

\begin{figure}[htb]
 \centering \epsfxsize=10cm \epsffile{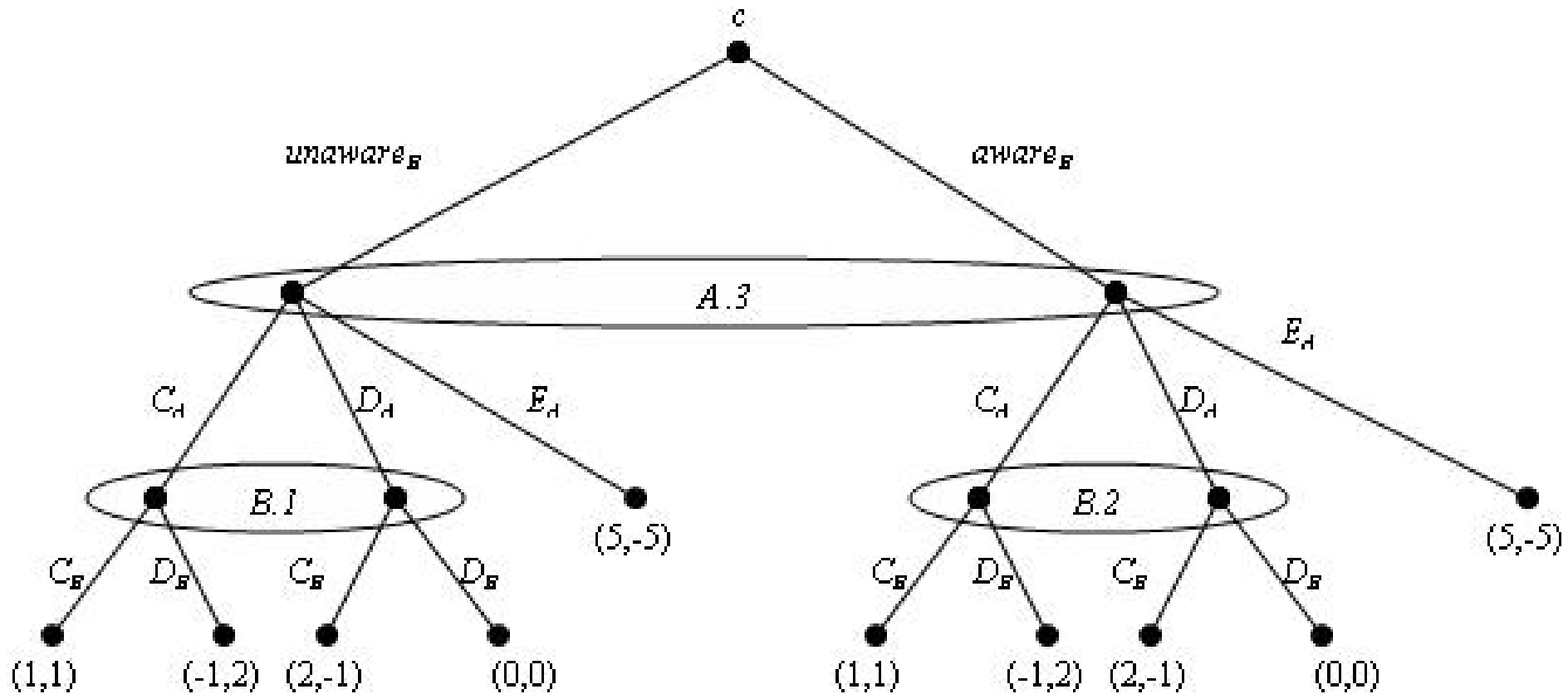}
\caption{$\Gamma^A$.} \label{chap2:fig:gammaA}
\end{figure}

\begin{figure}[htb]
 \centering \epsfxsize=3cm \epsffile{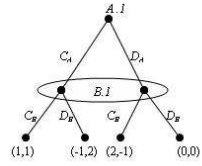}
\caption{$\Gamma^{B.1}$.} \label{chap2:fig:gammaB1}
\end{figure}

The game $\Gamma^{\nu}$ is the result of pasting together
$\Gamma^m$,
$\Gamma^A$, $\Gamma^{B.1}$, and $\Gamma^{B.2}$. There are 5 players:
$(A,\Gamma^A)$, $(A,\Gamma^{B.1})$, $(A,\Gamma^{B.2})$,
$(B,\Gamma^{B.1})$, and $(B,\Gamma^{B.2})$. $(A,\Gamma^{B.1})$ and
$(B,\Gamma^{B.1})$ are playing standard prisoner's dilemma and
therefore both should defect with probability 1; $(B,\Gamma^{B.1})$
must believe he is in the history where $(A,\Gamma^{B.1})$ defected
with probability 1. $(A,\Gamma^A)$ and $(A,\Gamma^{B.2})$ choose the
extra move $E_A$ with probability 1, since it gives $A$ a payoff of
5. The subtlety arises in the beliefs of $(B,\Gamma^{B.2})$ in the
generalized information set $(\Gamma^{B.2},\{C_A,D_A\})^*$, since
this generalized information set is reached with probability zero.
Note that
$(\Gamma^{B.2},\{C_A,D_A\})^*
=
\{\<\Gamma^m,C_A\>,$ $\<\Gamma^m,D_A\>$,
$\<\Gamma^A,\mbox{aware}_B,C_A\>$, $\<\Gamma^A,\mbox{aware}_B,D_A\>,
\<\Gamma^{B.2},C_A\>,$ $\<\Gamma^{B.2},D_A\>\}.$
By the definition of sequential equilibrium, player
$(B,\Gamma^{B.2})$ will have to consider a sequence of strategies
where all these histories are assigned positive probability.
Although in general this is not a problem, note that
$(B,\Gamma^{B.2})$ is meant to represent the type of player $B$ that
considers only histories in game $\Gamma^{B.2}$ possible. Thus,
intuitively, he should assign positive probability only to the
histories $\{\<\Gamma^{B.2},C_A\>,\<\Gamma^{B.2},D_A\>\}$.

To see how this leads to a problem, first note that there
is a sequential equilibrium of $\Gamma^{\nu}$ where
$(B,\Gamma^{B.2})$ believes with probability 1 that the true history
is $\<\Gamma^A,$aware$_B,C_A\>$,
$(A,\Gamma^{B.2})$ chooses $E_A$ with probability 1, and
$(B,\Gamma^{B.2})$ chooses $C_B$ with
probability 1.
It is rational for $(B,\Gamma^{B.2})$ to choose $C_B$
because $(B,\Gamma^{B.2})$ assigns probability 1 to the first move of nature
in $\Gamma^{\nu}$ be $\Gamma^A$.  Since his utility is 0 for every run
in $\Gamma^{\nu}$ whose first move is $\Gamma^A$, his expected utility is 0 no matter what move he
makes at the generalized information set, given his beliefs.

There is no reasonable definition of generalized sequential
equilibrium corresponding to this sequential equilibrium of
$\Gamma^{\nu}$. Player $B$ while moving at node $B.2$ would never
cooperate, since this is a strictly dominated strategy for him in
the game that he considers to be the actual game, namely
$\Gamma^{B.2}$.

The problem is that there is nothing in the definition of sequential
equilibrium that guarantees that the belief system of a sequential
equilibrium in $\Gamma^{\nu}$ assigns probability zero to histories
that players are unaware of in the game $\Gamma^*$ with awareness.
We want to define a modified notion of sequential
equilibrium for standard games that guarantees that the belief
system in $\Gamma^{\nu}$ associates each information set with a
probability distribution over a pre-specified subset of the
histories in the information, which consists only of the histories
in the information set player $i$ actually considers possible. In
this example, the pre-specified subset would be
$\{\<\Gamma^{B.2},C_A\>,\<\Gamma^{B.2},D_A\>\}$.
}
\section{CONDITIONAL SEQUENTIAL EQUILIBRIUM}
\label{chap2:sec:stronseq}

In the standard definition of sequential equilibrium for extensive
games, it is implicitly assumed that every player considers all
histories in his information set possible. This is evident from the
fact that if a strategy profile that is part of a sequential
equilibrium assigns positive probability to every move, then by the
consistency requirement the belief system also assigns positive
probability to every history of every information set of the game.
Therefore, this notion of equilibrium is not strong enough to
capture situations where a player is certain that some histories
in his information set will not occur.
The notion of {\em conditional sequential equilibrium}, which we now
define, is able to deal with such situations.
It generalizes sequential equilibrium: in a game where every player
considers every
history in his information set possible, the set of conditional
sequential equilibria and the set of sequential equilibria
coincide.

Given a standard extensive game $\Gamma$, define a {\em possibility
system} $\K$ on $\Gamma$ to be a function that determines for every
information set $I$ a nonempty subset of $I$ consisting of the
histories in $I$ that the player moving at $I$ considers possible.
We assume that $\K$ is common knowledge among players of the game,
so that every player understands what histories are considered
possible by everyone else in the game. If $I$ is an $i$-information
set, intuitively $i$ should be indifferent among all runs that go
through histories in $I-\K(I)$, since $i$ believes that those runs
will not occur and every other player knows that. Thus, for a given
$\Gamma$, a possibility system $\K$ must satisfy the
following requirement: if $z$ and $z'$ are two runs going through
histories in $I-\K(I)$ and $I$ is an $i$-information set, then
$u_i(z)=u_i(z')$.

Given a pair $(\Gamma,\K)$, a {\em $\K$-assessment} is a pair
$(\vec{\sigma},\mu)$, where $\vec{\sigma}$ is a strategy profile of
$\Gamma$, and $\mu$ is a {\em restricted belief system}, i.e., a
function that determines for every information set $I$ of $\Gamma$ a
probability $\mu_I$ over the histories in $\K(I)$. Intuitively, if
$I$ is an information set for player $i$, $\mu_I$ is $i$'s
subjective assessment of the relative likelihood of the histories
player $i$ considers possible while moving at $I$, namely $\K(I)$.
As in the definition of sequential equilibrium, a $\K$-assessment $(\vec{\sigma},\mu)$ is
a {\em conditional sequential equilibrium with respect to $\K$} if
(a) at every information set where a player moves he chooses a best
response given the beliefs he has about the histories that he
considers possible in that information set and the strategies of
other players, and (b) his restricted beliefs must be consistent
with the strategy profile being played and the possibility system,
in the sense that they are calculated by conditioning the
probability distribution induced by the strategy profile over the
histories considered possible on the information set.
\commentout{
More formally,
$(\vec{\sigma},\mu)$ must satisfy the following two properties:
\begin{itemize}
\item {\em Sequential rationality:} For every information set $I$,
player $i$, and behavioral strategy $\sigma$ for player
$i$,
$$\EU_i((\vec{\sigma},\mu)\mid I)\geq
\EU_i(((\vec{\sigma}_{-i},\sigma),\mu)\mid I),$$
where $\EU_i((\vec{\sigma},\mu)\mid I)=\sum_{h\in \K(I)}\sum_{z\in
Z}\mu_I(h)\Pr_{\vec{\sigma}}(z\mid h)u_i(z)$

\item {\em Consistency:}
If, for every information set $I$, $\sum_{h\in
\K(I)}\Pr_{\vec{\sigma}}(h)>0$, then for every $h\in \K(I)$
$$\mu_I(h)=\frac{\Pr_{\vec{\sigma}}(h)}{\sum_{h'\in
\K(I)}\Pr_{\vec{\sigma}}(h')}.$$

Otherwise, there exists a sequence of conditional assessments
$(\vec{\sigma}^i,\mu^i)$ such that every player chooses all of his
moves with positive probability, $\mu^i$ is consistent with
$\vec{\sigma}^i$ and $\K$ in the above sense, and
$(\vec{\sigma}^i,\mu^i)$ converges pointwise to
$(\vec{\sigma},\mu)$.
\end{itemize}
}
Formally, the definition of $(\vec{\sigma},\mu)$ being a conditional
sequential equilibrium is identical to that of sequential equilibrium,
except that the summation in the definition of $\EU_i(\vec{\sigma},\mu)
\mid I)$ and $\mu_I(h)$ is taken over histories in $\K(I)$ rather than
histories in $I$.
It is immediate that
if $\K(I)=I$ for every information set $I$ of the
game, then the set of conditional sequential equilibria with respect
to $\K$ coincides with the set of sequential equilibria. The next
theorem shows that set of conditional sequential equilibria for a
large class of extensive games that includes $\Gamma^{\nu}$ is
nonempty.

\thm \label{chap2:thm:conditional1} Let $\Gamma$ be an extensive
game with perfect recall and countably many players such that (a)
each player has only finitely many pure strategies and (b) each
player's payoff depends only on the strategy of finitely many other
players.
Let $\K$ be an arbitrary possibility system. Then there exists at
least one $\K$-assessment that is a
conditional sequential equilibrium of $\Gamma$ with respect to $\K$.
\ethm

We now prove that every game  of awareness has a generalized sequential
equilibrium
by defining a possibility system
$\K$ on $\Gamma^{\nu}$ and showing that there is a one-to-one
correspondence between the set of conditional sequential equilibria
of $\Gamma^{\nu}$
with respect to $\K$
and the set of generalized sequential equilibria
of $\Gamma^*$.

\thm \label{chap2:thm:conditional2} For all probability measures
$\nu$ on $\G$, if $\nu$ gives positive probability to all games in
$\G$, and $\K(\<\Gamma',I\>^*)=\{\<\Gamma',h\>:h\in I\}$ for every
information set $\<\Gamma',I\>^*$ of $\Gamma^{\nu}$, then
$(\vec{\sigma}',\mu')$ is a generalized sequential equilibrium of
$\Gamma^*$ iff $(\vec{\sigma},\mu)$ is a conditional sequential
equilibrium of $\Gamma^{\nu}$ with respect to $\K$, where
$\sigma_{i,\Gamma'}(\<\Gamma^h\>\cdot
h')=\sigma'_{i,\Gamma'}(\Gamma^h,h')$ and $\mu'_{\Gamma',
I}=\mu_{\<\Gamma',I\>^*}$. \ethm

Since $\Gamma^{\nu}$ satisfies all the conditions of
Theorem~\ref{chap2:thm:conditional1}, it easily follows from
Theorems~\ref{chap2:thm:conditional1} and
\ref{chap2:thm:conditional2} that every game with awareness has at
least one generalized sequential equilibrium.

Although it is not true that every conditional sequential
equilibrium is also a sequential equilibrium of an arbitrary game,
the next theorem shows there is a close connection between these
notions of equilibrium.
If $(\vec{\sigma},\mu)$ is a conditional sequential equilibrium with
respect to some possibility system $\K$, then there exists a
belief system $\mu'$ such that $(\vec{\sigma},\mu')$ is a sequential
equilibrium.

\thm \label{chap2:thm:conn} For every extensive game $\Gamma$ with
countably many players where each player has finitely many pure
strategies and for every possibility system $\K$,
if $(\vec{\sigma},\mu)$ is a conditional sequential
equilibrium of $\Gamma$ with respect to $\K$, then there exists a
belief system $\mu'$ such that $(\vec{\sigma}, \mu')$ is a
sequential equilibrium of $\Gamma$. \ethm

\section{RATIONALIZABILITY AND GENERALIZED NASH EQUILIBRIUM}
\label{chap2:sec:grat}

In this section, we analyze the relationship between the notions of
rationalizability and generalized Nash equilibrium, providing some
more intuition about the latter.

The usual justification for Nash equilibrium is that a player's
strategy must be a best response to the strategies selected by other
players in the equilibrium, because he can deduce what those
strategies are. However, in most strategic situations, it is not the
case that a player can deduce the strategies used by other players.
Since every player tries to maximize his expected payoff and this is
common knowledge, the best that a player can hope to do is to deduce
a set of reasonable strategies for the other players. Here, we take
a ``reasonable strategy'' to be a best response to some reasonable
beliefs a player might hold about the strategy profile being played.
This is the intuition that the {\em rationalizability} solution
concept tries to capture. Even though a notion of rationalizability
for extensive-form games was proposed by Pearce \citeyear{Pearce84},
rationalizability is more widely applied in
normal-form games. In
this section, we explore the relationship between rationalizability
and generalized Nash equilibrium in games with awareness where in
the underlying game each player moves only once, and these moves are
made simultaneously (or, equivalently, a player does not know the
moves made by other players before making his own move).
We show that, given an underlying game $\Gamma$ satisfying this
requirement, a pure strategy profile contains only rationalizable
strategies iff it is the strategy profile used by the players in the
modeler's game in some (pure) generalized Nash equilibrium of a game
$\Gamma^*$ with awareness. If we think of rationalizability as
characterizing ``best response to your beliefs'' and Nash
equilibrium characterizing ``best response to what is actually
played'', then this result shows that in the framework of games with
awareness, since the game is not common knowledge, the line between
these two notions is somewhat blurred.

We start by reviewing the notion of rationalizability for standard
normal-form games. Let $\C_i$ be the set of available pure
strategies for player $i$; $\C=\times_{i\in N}\C_i$ is thus the set
of pure strategy profiles. Let $\Delta(M)$ denote the set of all
probability distributions on $M$. Suppose that each player $i$ is
rational and is commonly known to choose a strategy from a subset
$\D_i$ of $\C_i$. Let $\D_{-i}=\times_{j\ne i}\D_i$ and
\begin{eqnarray}
& & B(\D_{-i})=\{argmax_{s_i\in\C_i}\EU_i((s_i,\pi(\D_{-i}))):\nonumber\\
& & \mbox{ for
some }\pi\in\Delta(\D_{-i})\}; \nonumber
\end{eqnarray}
that is, $B(\D_{-i})$ consists of the strategies in $\C_i$ that are
best responses to some belief that player $i$ could have about the
strategies other players are using.

The set $\cS=\times_{i\in N}\cS_i$ of {\em correlated rationalizable
strategies } is characterized by the following two properties: (a)
for all $i \in N$, $\cS_i\subseteq B(\cS_{-i})$ and (b) $\cS$ is the
largest set satisfying condition (a), in the sense that, for every
set of strategy profiles $\D$ satisfying (a), we have that
$\D\subseteq \cS$.
It is not hard to show that for every player $i$, $\cS_i=
B(\cS_{-i})$. A strategy $s_i\in \cS_i$ is called a {\em correlated
rationalizable strategy for player $i$}.%
\footnote{From now on, we use $\vec{s}=(s_1,\ldots,s_n)$ to denote
pure strategy profiles, and will continue to use $\vec{\sigma}$ for
possibly nonpure strategy profiles.}
\footnote{In the literature, it is often assumed that each player
chooses his strategy
independently of the others and that this is common knowledge.  If we
make this assumption,
we get a somewhat stronger solution concept (at least, if $|N| \ge 3$),
which we call \emph{uncorrelated rationalizability}.  Essentially the
same results as we prove here for correlated rationalizability hold for
uncorrelated rationalizability; we omit further details here.}
\commentout{
Formally,
suppose that each player is rational and is commonly known to choose
a strategy from a subset $\D_i$ of $\C_i$. Let $\D_{-i}=\times_{j\ne
i}\D_i$ and
$$O(\D_{-i})=\{argmax_{s_i\in\C_i}\EU_i((s_i,\times_{j\in
N-\{i\}}\pi_j(\D_{j}))):\mbox{ for some }\pi_j\in\Delta(\D_j)\};$$
that is, $O(\D_{-i})$ consists of the strategies in $\C_i$ that are
best responses to some belief that player $i$ could have about the
strategies other players are using.
The set $\cS^u=\times_{i\in N}\cS_i^u$ of {\em uncorrelated
rationalizable
strategies } is characterized by the following two properties: (a)
for all $i \in N$, $\cS_i^u\subseteq O(\cS_{-i}^u)$ and (b) $\cS^u$
is the largest set satisfying condition (a), in the sense that, for
every set of strategy profiles $\D$ satisfying (a), we have that
$\D\subseteq \cS^u$.
Again, it is not hard to show that for every player $i$, $\cS^u_i=
O(\cS_{-i}^u)$. A strategy $s_i^u\in \cS_i^u$ is called an {\em
uncorrelated rationalizable strategy for player $i$}.
}
It turns out that we can construct $\cS$ by the following iterative
procedure. Let $\C_i^{0}=\C_i$ for all $i\in N$. Define
$\C_i^{j}=B(\C_{-i}^{j-1})$ for $j\geq 1$. Since there are finitely
many strategies it is easy to see that
there exists a finite $k$ such that $\C_i^j=\C_i^k$
for all $j\geq k$.
It can
be shown that $\cS_i=\lim_{j\rightarrow\infty} \C_i^j=\C_i^k$.
\commentout{The set of
uncorrelated rationalizable strategies
can be constructed in a similar way
(of course, replacing $B(\cdot)$ by $O(\cdot)$).}
It is also easy to see that if $\vec{\sigma}$ is a (behavioral) Nash
equilibrium, then every pure strategy
that is played with positive probability according to $\vec{\sigma}$
is rationalizable (where the probability with which a pure strategy
is played according to $\vec{\sigma}$ is the product of the
probability of each of its moves according to $\vec{\sigma}$).

We now explore the relationship between rationalizability in an
underlying game $\Gamma$ in normal-form and generalized Nash
equilibrium in a special class of games with awareness based on
$\Gamma$.
Given a standard game $\Gamma$ and a pure strategy profile $\vec{s}$
consisting of rationalizable strategies, we define a game with
awareness $\Gamma^*(\vec{s}) = (\G,\Gamma^m,\F)$ such that (a) there
exists a generalized Nash equilibrium of $\Gamma^*(\vec{s})$, where
$s_i$ is the strategy followed by player $i$ in $\Gamma^m$, and (b)
every local strategy in every pure generalized Nash equilibrium of
$\Gamma^*(\vec{s})$ is rationalizable in $\Gamma$.
To understand the intuition behind the construction, note that if
$s_i$ is a pure correlated rationalizable strategy of player $i$ in
$\Gamma$, then $s_i$ must
be a best response to some probability distribution $\pi^{s_i}$ over
the set $\cS_{-i}$ of pure correlated rationalizable strategies of
$i$'s opponents.
The idea will be to include a game $\Gamma^{s_i}$ in $\G$ that captures
the beliefs that make $s_i$ a best response.
Let $\vec{s}^{\,1}, \ldots, \vec{s}^{\,m}$ be the strategy profiles
in $\cS_{-i}$ that get positive probability according to
$\pi^{s_i}$. (There are only finitely many, since $\cS_{-i}$
consists of only pure strategies.)
Let $\Gamma^{s_i}$ be the game where nature initially makes one of
$m$ moves, say $c_1, \ldots, c_m$ (one corresponding to each
strategy that
gets positive probability according to $\pi^{s_i}$), where the
probability of move $c_j$ is $\pi^{s_i}(\vec{s}^{\,j})$. After
nature's choice a copy of $\Gamma$ is played.
All the
histories in $\Gamma^{s_i}$ in which player $i$ is about to move are in
the same information set of player $i$; that is, player $i$ does not
know nature's move.  However, all the other players know nature's move.
Finally, all players are aware of all runs of
$\Gamma$ at every history in $\Gamma^{s_i}$.
Note that if $h$ is a history where player $i$ thinks game
$\Gamma^{s_i}$ is the actual game, and believes that other players
will play $\vec{s}_{-i}^{\,j}$ after nature's move
$c_j$, then player $i$ believes that $s_i$ is a best response at
$h$.
Given a pure strategy profile $\vec{s}$ of the game $\Gamma$, let
$\Gamma^*(\vec{s}) = (\G,\Gamma^m,\F)$ be the following game with
awareness:
\begin{itemize}
\item $\Gamma^m = (\Gamma,\{A_i:i\in N\})$, where for every player $i$
and every history $h\in H_i^m$, $A_i(h)=H$ (the set of all histories in $\Gamma$);

\item $\G=\{\Gamma^m\}\union \{\Gamma^{s'_i}:s'_i\in \cS_i, i\in N\}$;

\item for an augmented game in $\Gamma^+\in (\G-\{\Gamma^m\})$ and
a history $h$ of $\Gamma^+$ of the form $\<s'_{-i}\>\cdot h'$,
\begin{itemize}
\item if $P^+(h)=i$, then $\F(\Gamma^+,h)=(\Gamma^+,I)$ where $I$ is the information set containing $h$;
\item if $P^+(h)=j\in N-\{i\}$ and $s'_j$ is the strategy of player $j$ specified by $s'_{-i}$, then $\F(\Gamma^+,h)=(\Gamma^{s'_j},I)$, where $I$ is the unique
$j$-information set in game $\Gamma^{s'_j}$;
\end{itemize}

\item for $h\in H_i^m$, $\F(\Gamma^m,h)=(\Gamma^{s_i},I)$, where $I$ is the unique
$i$-information set in game $\Gamma^{s_i}$.
\end{itemize}

The intuition is that if $\vec{s}$ is a strategy profile such that,
for all $i \in N$, $s_i$ is a rationalizable strategy for player $i$
in $\Gamma$, then
at the (unique) $i$-information set of $\Gamma^m$, $i$ considers the
actual game to be $\Gamma^{s_i}$. For this particular game with
awareness,
there exists a generalized Nash equilibrium where the strategy for
each player $i$ in the modeler's game is $s_i$. Conversely, only
rationalizable strategies are used in any
pure
generalized Nash equilibrium of
$\Gamma^*(\vec{s})$.
There is only one small problem with this intuition: strategies in
$\Gamma$ and local strategies for augmented games in
$\Gamma^*(\vec{s})$ are defined over different objects. The former
are defined over information sets of the underlying game $\Gamma$
and the latter are defined over generalized information sets of
$\Gamma^*(s)$.
Fortunately, this problem is easy to deal with: we can identify a
local
strategy in $\Gamma^*(\vec{s})$ with a strategy in $\Gamma$ in the
obvious way.
By definition of $\Gamma^*(\vec{s})$, for every player $i$ and
augmented game $\Gamma'\in \G_i$, the domain of the local strategy
$\sigma_{i,\Gamma'}$ consists of a unique generalized information
set. We denote this information set by $I_{i,\Gamma'}$. For each
local strategy $\sigma_{i,\Gamma'}$ of $\Gamma^*(\vec{s})$, we
associate the strategy $\underline{\sigma}_{i,\Gamma'}$ in the
underlying game $\Gamma$ such that
$\sigma_{i,\Gamma'}(I_{i,\Gamma'})=\underline{\sigma}_{i,\Gamma'}(I)$,
where $I$ is the unique $i$-information set in $\Gamma$.

The following theorem summarizes the relationship between correlated
rationalizable strategies in $\Gamma$ and generalized Nash
equilibrium of games with awareness.

\thm \label{chap2:rat_gNash} If $\Gamma$ is a standard normal-form
game and $\vec{s}$ is a (pure) strategy profile such that for all
$i\in N$, $s_i$ is a correlated rationalizable strategy of player
$i$ in $\Gamma$, then
\begin{itemize}
\item[(i)] there is a (pure) generalized Nash equilibrium $\vec{s}^{\,*}$ of
$\Gamma^*(\vec{s})$ such that for every player $i$,
$\underline{s}^*_{i,\Gamma^{s_i}} =s_i$;

\item[(ii)] for every (pure) generalized Nash equilibrium $\vec{s}^{\,*}$ of
$\Gamma^*(\vec{s})$, for every local strategy $s^*_{i,\Gamma'}$ for
every player $i$ in $\vec{s}^{\,*}$,
the strategy $\underline{s}^*_{i,\Gamma'}$ is correlated
rationalizable for player $i$ in $\Gamma$.
\end{itemize}
\ethm

\commentout{A theorem analogous to Theorem~\ref{chap2:rat_gNash} also
holds for the uncorrelated rationalizability case; we leave details to
the reader.}

Note that Theorem~\ref{chap2:rat_gNash} does not imply that for a
fixed game with awareness, (pure) generalized Nash equilibrium and
generalized rationalizability coincide. These notions are
incomparable for standard extensive games (cf.~%
\cite{Bat97,Pearce84}),
so the corresponding generalized notions are
incomparable when applied to the canonical representation of a
standard game as a game with awareness. If we restrict the
underlying game to be
in normal form, it can be shown that, just as in standard games,
every strategy in a
pure
generalized Nash equilibrium is (generalized) rationalizable.
Since rationalizability is usually defined for
pure strategies in
the literature \cite{Myerson,OR94}, we
focused on that case here. But it is not hard to show that an
analogue of Theorem~\ref{chap2:rat_gNash} holds for behavioral
rationalizable strategies as well.

\commentout{
\section{DISCUSSION}\label{sec:discussion}

Our focus in this paper has been on refinements of generalized Nash
equilibrium in games with awareness.
It is worth reconsidering here the conceptual basis for these solution
concepts in the presence of awareness.%
\footnote{We thank Aviad Heifetz and an anonymous referee for raising
some of these issues.}
As we noted, equilibrium refinements are used in standard games to
eliminate some ``undesirable'' or ``unreasonable'' equilibria.
Arguably, unreasonable equilibria pose an even deeper problem with
unaware players.
For example,
one standard interpretation of a Nash equilibrium is that a player
chooses his strategy at the beginning of the game, and then does not
change it because he has no motivation for doing so (since his payoff
is no higher when he changes strategies).  But this interpretation is
suspect in extensive-form games when a player makes a move that
takes the game off the equilibrium path.  It may seem unreasonable for a
player to then play
the move called for by his strategy (even if the strategy is part of a
Nash equilibrium).
A threat to blow up the world if I catch you
cheating in a game may be part of a Nash equilibrium, and does not cause
problems if in fact no one cheats, but it hardly seems credible if
someone does cheat.

One way to justify the existence of incredible
threats off the equilibrium path in a Nash equilibrium is to view the
player as
choosing
a computer program
that will play the game for him,
and then leaving.
Since the program is not changed once it is set in motion, threats about
moves that will be made at information sets off the equilibrium path
become more credible.
However, in a game with awareness, a player cannot write a program to
play the whole game at the beginning of the game because, when his level of
awareness changes, he realizes that there are moves available to him
that he was not aware of at the beginning of the game.
He thus must write a new program that takes this into account.
But this means we cannot sidestep the problem of incredible threats by
appealing to the use of a pre-programmed computer to play a strategy.
Once we allow a player to change his program, threats that were made
credible because the program could not be rewritten become incredible
again.  Thus, the consideration of equilibrium refinements such as
sequential equilibrium, which block incredible threats, becomes even
more pertinent with awareness.
}

Moving up a level, we might ask more generally for the appropriate
interpretation of Nash equilibrium in games with awareness.
In standard games with a unique Nash equilibrium, we could perhaps argue
that rational players will play their component of the equilibrium,
since they can compute it and realize that it is the only stable
strategy.  In games with several Nash equilibria, perhaps one can be
singled out as most salient, or some can be eliminated by using
refinements of Nash equilibria.

To some extent, these considerations apply in games with awareness as
well. If there is a unique generalized Nash equilibrium, although a
player cannot necessarily compute the whole equilibrium (for example, if
it involves moves that he is not aware of), he can compute that part of
the equilibrium that is within the scope of his awareness.  Thus, this
argument for playing a Nash equilibrium lifts from standard games to
games with awareness.  However, other arguments do not lift so well.
For example, in standard games, one argument for Nash equilibrium is
that, over time, players will learn to play a Nash equilibrium, for
example, by playing a best response to their current beliefs.
\fullv{(However, this is not true in general \cite{Nachbar97,Nachbar05}.}
This argument will not work in the presence of awareness, since playing
the game repeatedly can make players aware of moves or of other players
awareness, and thus effectively change the game altogether.

Another way of interpreting Nash equilibrium in standard games
is in terms of evolutionary game theory.
This approach works with awareness as well.
Suppose
that we have populations consisting of each awareness type of each player, and
that at each time step we draw without
replacement one individual of each of these populations and let them
play the game once. If the sample individuals are playing an equilibrium
strategy, they do not have incentive to deviate unilaterally given their
beliefs that the other players will continue to follow the equilibrium
strategies.

\commentout{
In \cite{HR06}, we discussed modifications of games with awareness to
include situations where players may be aware of their own unawareness
and also games where players may have no common knowledge of what the
underlying game is, for example, they may disagree on what the payoffs
or the information sets are. Therefore with these models, we can capture
a situation where players may think that they move alone but find out
later that in fact they moved simultaneously with another player.
It is not hard to show that the results of this paper can be extended to these
situations as well.
}

\commentout{
Other issues arise when considering sequential equilibrium in games with
awareness.  For example,
in a standard game, when a player reaches a history that is not on the
equilibrium path, he must believe that his opponent made a mistake.
However, in games with awareness,
a player may become aware of her own unawareness and, as a result,
switch strategies.
In the definition of sequential equilibrium in standard games, play off
the equilibrium path is dealt with by viewing it as the limit of ``small
mistakes'' (i.e., small deviations from the equilibrium strategy).
Given that there are alternative ways of dealing with mistakes in games
with awareness, perhaps other approaches for dealing with
off-equilibrium play might be more appropriate.
While other ways of dealing with mistakes may well prove interesting, we
would argue that our generalization of sequential equilibrium can be
motivated the same way as in standard games.
Roughly speaking, for us, how a player´s awareness level changes over
time is not part of the equilibrium concept, but is given as part of the
description of the game.}

\commentout{
More generally,
we have focused here on generalizing solution concepts that have proved
useful in standard games, where there is no lack of awareness. The
discussion above suggests that introducing awareness allows us to consider
other solution concepts.
For example,
Ozbay \citeyear{Ozbay06} proposes an approach where a player's
beliefs about the probability of  revealed moves of nature, that the
player was initially unaware of, are formed as part of the equilibrium
definition.
We hope to explore the issue of which solution concepts are most
appropriate in games with awareness in future work.}

\commentout{
Li \citeyear{LI06b} has also provided a model of unawareness in
extensive games, based on her earlier work on modeling
unawareness \cite{LI06,LI06a}. Although her representation of a game
with unawareness is quite similar to ours, her notion of equilibrium is
a generalization of Nash equilibrium for standard games, but it is
different from the one we proposed in \cite{HR06}.
}

\fullv{
\section{CONCLUSIONS}
\label{chap2:sec:conc}

In this paper, we further developed the framework of games with
awareness by analyzing how to generalize sequential equilibrium to such
games.
Other solution concepts can be generalized
in a similar way.  Although we
have not checked all the details for all solution concepts, we believe that
techniques like those used in our earlier paper to prove existence of
generalized Nash
equilibrium
and
ones similar to those used in this
paper for generalized sequential equilibrium will be useful for proving
the existence of other generalized solution concepts.
For example, consider the notion of {\em (trembling hand)
perfect equilibrium} \cite{Selten75}.
in
normal-form games.
A strategy profile $\vec{\sigma}$ is a perfect
equilibrium if there exists some sequence of strategies
$(\vec{\sigma}^k)_{k=0}^{\infty}$, each assigning positive
probability to every available move, that converges pointwise to
$\vec{\sigma}$
such that for each player $i$, the strategy $\sigma_i$ is a best
response to $\vec{\sigma}^k_{-i}$ for all $k$.
The
definition of {\em generalized perfect equilibrium} in games with
awareness is the same as in standard games, except that we use
generalized strategies rather than strategies, and require that for
every local strategy $\sigma_{i,\Gamma'}$ of every player $i$,
$\sigma_{i,\Gamma'}$ is a best response to
$\vec{\sigma}^k_{-(i,\Gamma')}$ in game $\Gamma'$ for all $k$.
To
prove that every game with awareness in normal form has a generalized
perfect equilibrium,
we prove an analogue of Theorem~3.1(b) in \cite{HR06}, giving a
correspondence between the set of generalized perfect equilibria of
$\Gamma^*$ and the set of perfect equilibria of $\Gamma^{\nu}$.
The existence of a generalized perfect equilibrium follows from the
existence of a perfect equilibrium in $\Gamma^{\nu}$; the existence
of a perfect equilibrium in $\Gamma^{\nu}$ follows from
Lemma~\ref{chap2:lem:exisperf}.

\commentout{As a byproduct of our proof of existence of
a generalized sequential equilibrium in each game with awareness, we proposed
a concept of
conditional sequential equilibrium for standard games. This solution
is more appropriate than standard sequential equilibrium if there
are histories in an information set that, even though
indistinguishable from a player's point of view, are not considered
possible. We showed how to construct a standard game given a game
with awareness such that there is a one-to-one correspondence
between the set of generalized sequential equilibria of the game
with awareness and the set of conditional sequential equilibria of
the standard game. Roughly speaking, this result shows that a game
with awareness is equivalent to a standard game where there are
multiple versions of a player and it is possible that a player does
not consider the actual history possible if it involves moves he is
unaware of.}

In our earlier work, we showed that our definitions could be extended in
a straightforward way to games with awareness of unawareness;
that is, games where one player might be aware that there are
moves that another player (or even she herself) might be able to
make, although she is not aware of what they are.
Such awareness of unawareness can be quite relevant in practice.
We captured the fact that player $i$ is aware that,
at a node $h$ in the game tree, there is a move that $j$ can make
she ($i$) is not aware of was by having $i$'s subjective representation of
the game include a ``virtual'' move for $j$ at node $h$.
Since $i$
does not understand perfectly what can happen after this move, the
payoffs associated with runs that follow a virtual move represent what
player $i$ believes will happen if this run is played and
may bear no relationship to the actual payoffs in the underlying
game.
We showed that a generalized Nash equilibrium exists in games with
awareness of unawareness.
It is straightforward to define generalized sequential
equilibrium for those games
and to prove its existence
using the techniques of this paper;
we leave details to the reader.

\commentout{
We also provided a further insight into the notion of generalized
Nash equilibrium by analyzing its relationship with the notion of
rationalizability for standard games. We showed that, in a sense,
generalized Nash equilibrium can be viewed as a generalization of
rationalizability. In particular, this shows that, unlike in the
standard case where Nash equilibrium is characterized by every
player best responding to the actual strategies played by their
opponents, in games with awareness (or, more generally, in games
with lack of common knowledge) a generalized Nash equilibrium is
characterized by every player best responding to the strategy they
believe their opponents are playing.}

\commentout{
We have focused here on generalizing solution concepts that have proved
useful in games where there is no lack of awareness.  It may well be the
case that, in games with (lack of) awareness, other solution concepts
may also be appropriate.  We hope to explore the issue of which solution
concepts are most appropriate in games with awareness in future work.}

}

We have focused here on generalizing solution concepts that have proved
useful in standard games, where there is no lack of awareness. Introducing awareness allows us to consider
other solution concepts.
For example, Ozbay \citeyear{Ozbay06} proposes an approach where a player's
beliefs about the probability of  revealed moves of nature, that the
player was initially unaware of, are formed as part of the equilibrium
definition. We hope to explore the issue of which solution concepts are most
appropriate in games with awareness in future work.

\subsubsection*{Acknowledgments}
We thank Yossi Feinberg and Aviad Heifetz for useful
discussions on awareness.
We also thank Larissa S. Barreto for spotting some typos in a previous
version of this paper.
This work was supported in part by NSF under grants CTC-0208535, ITR-0325453, and
IIS-0534064, by ONR under grants N00014-00-1-03-41 and
N00014-01-10-511, and by the DoD Multidisciplinary University
Research Initiative (MURI) program administered by the ONR under
grant N00014-01-1-0795.
Some of this work was done while the first author was at the School
of Electrical and Computer Engineering at Cornell University,
U.S.A., supported in part
by a scholarship from the Brazilian Government through the Conselho
Nacional de Desenvolvimento Cient\'ifico e Tecnol\'ogico (CNPq).
\bibliographystyle{chicagor}
\bibliography{z,joe}

\appendix

\fullv{\section{PROOF OF THEOREMS}}

\shortv{\section{THE DEFINITION OF $\Gamma^\nu$}
We now give the formal definition of the standard game $\Gamma^\nu$ used
to show that there is a generalized sequential equilibrium.
Given a game $\Gamma^* = (\G, \Gamma^m, \F)$ with awareness, let
$\nu$ be a probability on $\G$ that assigns each game in $\G$
positive probability.  (Here is where we use the fact that $\G$
is countable.)
For each $\Gamma'\in \G$, let $\lfloor
H^{\Gamma'}\rfloor=\{h\in H^{\Gamma'}:$ for every prefix $h_1\cdot
\<m\>$ of $h$, if $P'(h_1)=i \in N$ and
$\F(\Gamma',h_1)=(\Gamma'',I)$, then for all $h_2\in I$,
$h_2\cdot\<m\>\in \H''\}$.
The histories in $\lfloor
H^{\Gamma'}\rfloor$ are the ones that can actually be played according to the
players' awareness levels.

Let $\Gamma^\nu$ be the standard game such that
\begin{itemize}
\item $N^\nu = \{(i,\Gamma'):\Gamma'\in\G_i\}$;

\item $\M^\nu = \G\union_{\Gamma'\in\G}\lfloor M^{\Gamma'}\rfloor$, where
$\lfloor M^{\Gamma'}\rfloor$ is the set of moves that occur in
$\lfloor H^{\Gamma'}\rfloor$;

\item $\H^\nu = \<\, \>\union\{\<\Gamma'\>\cdot h:\Gamma'\in\G, h\in
\lfloor H^{\Gamma'}\rfloor\}$;

\item $P^\nu(\<\, \>)=c$, and
$$P^\nu(\<\Gamma^h\>\cdot h') =
\left\{ \begin{array}{ll} (i,\Gamma^{h'}) & \mbox{if $P^h(h') = i \in
N$ and }\\
\ & \F(\Gamma^h,h')=(\Gamma^{h'}, \cdot),\\
c & \mbox{if $P^h(h') = c$;}\end{array} \right.$$

\item $f_c^\nu(\Gamma'|\<\, \>)= \nu(\Gamma')$ and
$f_c^\nu(\cdot|\<\Gamma^h\>\cdot h') = f_c^h(\cdot|h')$ if $P^h(h')
= c$;

\item $\I^\nu_{i,\Gamma'}$ is a partition of $\H^{\nu}_{i,\Gamma'}$ where two histories $\<\Gamma^1\>\cdot h^1$ and $\<\Gamma^1\>\cdot h^1$
are in the same information set $\<\Gamma',I\>^*$ iff
$(\Gamma^{1},h^1)$ and $(\Gamma^{2},h^2)$ are in the same
generalized information set $(\Gamma',I)^*$;

\item $u_{i,\Gamma'}^\nu(\<\Gamma^h\>\cdot z)= \left\{
\begin{array}{ll}
u_i^h(z) &\mbox{if $\Gamma^h = \Gamma',$}\\
0 &\mbox{if $\Gamma^h \ne \Gamma'.$}\end{array} \right.$
\end{itemize}
}
\fullv{

\subsection{PROOF OF THEOREMS~\ref{chap2:thm:conditional1},
\ref{chap2:thm:conditional2}, AND \ref{chap2:thm:conn}}

\othm{chap2:thm:conditional1} Let $\Gamma$ be an extensive game with
perfect recall and countably many players such that (a) each player
has only finitely many pure strategies and (b) each player's payoff
depends only on the strategy of finitely many other players.
Let $\K$ be an arbitrary possibility system. Then there exists at
least one $\K$-assessment that is a conditional sequential
equilibrium of $\Gamma$ with respect to $\K$. \eothm

\prf We use the same ideas that are used to prove existence of
standard sequential equilibrium, following closely the presentation
in \cite{Myerson}. The proof goes as follows. Given $\Gamma$, let
$\Gamma_M$ be the {\em multiagent representation} of $\Gamma$ in
normal form. We prove (Lemma~\ref{chap2:lem:perfseq}) that for every
{\em perfect equilibrium} $\sigma$ of $\Gamma_M$ and every
possibility system $\K$, there exists a restricted belief system
$\mu$ such that $(\sigma,\mu)$ is a conditional sequential
equilibrium of $\Gamma$ with respect to $\K$. Then we show
(Lemma~\ref{chap2:lem:exisperf}) that for $\Gamma$ satisfying the
hypothesis of the theorem, $\Gamma_M$ has at least one perfect
equilibrium.

We now review the relevant definitions. A {\em normal-form game} is
a tuple $(N,$ $\times_{i\in N}\C_i,\{u_i:i\in N\})$, where $N$ is
the set of players of the game, $\C_i$ is the collection of pure
strategies available for player $i$ in the game, and $u_i$ is a
payoff function that determines for each strategy profile in
$\times_{i\in N}\C_i$ the payoff for player $i$.

Given a standard extensive-form game
$\Gamma=(N,M,H,P,f_c,\{\I_i:i\in N\},\{u_i:i\in N\})$, let
$S^*=\union_{i\in N}\I_i$. Intuitively, we associate with each
$i$-information set $I\in \I_i$ a {\em temporary player} that has
$M(I)$ as its set possible strategies; $S^*$ is just the set of all
temporary players in $\Gamma$. For each temporary player $I$ we
associate a payoff function $v_I:\times_{I\in S^*}M(I)\rightarrow
\mathrm{R}$ such that if each temporary player $I$ chooses action
$a_I$, and $\sigma$ is the pure strategy profile for $\Gamma$ such
that for every $i\in N$ and $I\in \I_i$, $\sigma_i(I)=a_I$, then
$v_I(\times_{I\in S^*}a_I)=u_i(\sigma)$. The {\em multiagent
representation for $\Gamma$ in normal form} is the tuple
$(S^*,\times_{I\in S^*}M(I),\{v_I:I\in S^*\})$.

Given any countable set $B$, let $\Delta(B)$ be the set of all
probability measures over $B$, and let $\Delta^0(B)$ be the set of
all probability measures over $B$ whose support is all of $B$. Given
a game in normal form $\Gamma=(N,\times_{i\in N}\C_i,\{u_i:i\in
N\})$, a mixed strategy profile $\sigma\in \times_{i\in
N}\Delta(\C_i)$ is a {\em perfect equilibrium} of $\Gamma$ iff there
exists a sequence $(\hat{\sigma}^k)_{k=1}^{\infty}$ such that (a)
$\hat{\sigma}^k\in \times_{i\in N}\Delta^0(\C_i)$ for $k\geq 1$, (b)
$\hat{\sigma}^k$ converges pointwise to $\sigma$, and (c)
$\sigma_i\in argmax_{\tau_i\in
\Delta(\C_i)}\EU_i(\hat{\sigma}^k_{-i},\tau_i)$ for all $i\in N$.

The following lemmas are analogues of Theorems 5.1 and 5.2 in
\cite{Myerson}.

\lem \label{chap2:lem:perfseq} If $\Gamma_M$ is a multiagent
representation of $\Gamma$ in normal form, then for every perfect
equilibrium $\sigma$ of $\Gamma_M$ and every possibility system
$\K$, there exists a restricted belief system $\mu$ such that
$(\sigma,\mu)$ is a conditional sequential equilibrium of $\Gamma$
with respect to $\K$. \elem

\prf The proof is almost identical to that of
Theorem 5.1 in \cite{Myerson}. We focus on the necessary changes,
leaving the task of verifying that the rest of the proof goes without
change to the reader. Let $(\hat{\sigma}^k)_{k=1}^{\infty}\in \times_{I\in
S^*}\Delta(M(I))$ be a sequence of behavioral strategy profiles
satisfying conditions (a), (b), and (c) of the definition of perfect
equilibrium. For each $k$, define a belief system $\mu^k$ such that,
for each information set $I$, $\mu^k(I)$ is the probability over
histories in $\K(I)$ defined as
$$\mu^k_I(h)=\frac{\Pr_{\hat{\sigma}^k}(h)}{\sum_{h'\in \K(I)}\Pr_{\hat{\sigma}^k}(h')}.$$
If $\I$ is the set all information sets in $\Gamma_M$, for each $k$,
$\mu^k : \I \rightarrow [0,1]$.  Thus, $\mu^k \in [0,1]^{\I}$; and,
by Tychonoff's Theorem, $[0,1]^{\I}$ is compact.   Thus, there must
be a convergent subsequence of $\mu^1, \mu^2, \ldots$.   Suppose
that this subsequence converges to $\mu$. It is easy to see that
$\mu$ is consistent with $\sigma$ and $\K$.

Let $Z(I)$ denote the set of runs that do not contain any prefix in
$I$. Let $I$ be an arbitrary $i$-information set of $\Gamma^{\nu}$.
When agent $I\in S^*$ uses the randomized strategy $\rho_I\in
\Delta(M(I))$ against the strategies specified by $\hat{\sigma}^k$
for all other agents, his expected payoff is
$$\EU_I(\hat{\sigma}^k_{-I},\rho_I)=\sum_{h\in
I}\Pr_{(\hat{\sigma}^k_{-I},\rho_I)}(h)\EU_I(\hat{\sigma}^k_{-I},\rho_I\mid
h)+\sum_{z\in Z(I)}\Pr_{(\hat{\sigma}^k_{-I},\rho_I)}(z)u_i(z).$$
Note that for $h\in I$ or $h\in Z(I)$,
$\Pr_{(\hat{\sigma}^k_{-I},\rho_I)}(h)=\Pr_{\hat{\sigma}^k}(h)$, since
this probability is independent of the strategy used by player $I$.
Also note that for all $h\in I-\K(I)$,
$\EU_I(\hat{\sigma}^k_{-I},\rho_I\mid h)$ is independent of $\rho_I$.
Thus,
\begin{eqnarray}
& & \EU_I(\hat{\sigma}^k_{-I},\rho_I) \nonumber\\
& & =\sum_{h\in
\K(I)}\Pr_{\hat{\sigma}^k}(h)\EU_I(\hat{\sigma}^k_{-I},\rho_I\mid
h)+\sum_{z\in Z(I)}\Pr_{\hat{\sigma}^k}(z)u_i(z) + C'\nonumber \\
& & = (\sum_{h\in
\K(I)}\mu^k_I(h)\EU_I(\hat{\sigma}^k_{-I},\rho_I\mid h))(\sum_{h\in
\K(I)}\Pr_{\hat{\sigma}^k}(h)) +C'',\nonumber
\end{eqnarray}
where $C'$ and $C''$ are two constants independent of $\rho_I$.

The rest of the proof proceeds just as the proof of Theorem 5.1 in
\cite{Myerson}; we omit details here. \eprf

\lem \label{chap2:lem:exisperf} If $\Gamma$ is an extensive-form
game with perfect recall such that (a) there are at most countably
many players, (b) each player has only finitely many pure
strategies, and (c) the payoff of each player depends only on the
strategy of finitely many other players, then $\Gamma_M$ has at
least one perfect equilibrium. \elem

\prf The proof is almost identical to that of
Theorem 5.2 in \cite{Myerson}. Again, we focus on the necessary
changes, leaving it to the reader to verify that the rest of the
proof goes without change. We need to modify some of the arguments
since $\Gamma_M$ is not a finite game; since it may contain
countably many players. First, by the same argument used to prove
that $\Gamma^{\nu}$ has at least one Nash equilibrium in our earlier
work \cite{HR06},
we have that for any $\Gamma$ satisfying the hypothesis of the lemma,
$\Gamma_M$
has at least one Nash equilibrium.

Let $\C_i$ be the set of pure strategies available for player $i$ in
$\Gamma_M$.
\commentout{ We first need to show that if
$(\sigma^k)_{k=1}^{\infty}$ is a sequence of strategy profiles, then
we can find a subsequence that converges pointwise. We construct
such a subsequence as follows. Since, $\Delta(\C_1)$ is a compact
set, let $(\sigma^{1,j})_{j=1}^{\infty}$ be a subsequence of
$\sigma^k$ such that $\sigma^{1,j}_1$ converges pointwise to
$\sigma_1$. Define $(\sigma^{l,j})_{j=1}^{\infty}$ recursively by
taking it to be a subsequence of $(\sigma^{l-1,j})_{j=1}^{\infty}$
such that $\sigma^{l,j}_l$ converges pointwise to $\sigma_l$. The
sequence $\sigma^{j,j}$ provides a subsequence of $\sigma^k$ such
that it converges to $\sigma$ pointwise, as desired.}
By Tychonoff's Theorem $\times_{i\in
N}[0,1]^{\C_i}$ is compact. Since $\times_{i\in
N}\Delta(\C_i)$ is a closed subset of $\times_{i\in
N}[0,1]^{\C_i}$, it is also compact. All the remaining steps of the proof of
Theorem 5.2 in \cite{Myerson} apply here without change; we omit
the details. \eprf

The proof of Theorem~\ref{chap2:thm:conditional1} follows
immediately from Lemmas~\ref{chap2:lem:perfseq} and
\ref{chap2:lem:exisperf}.~\eprf

\othm{chap2:thm:conditional2} For all probability measures $\nu$ on
$\G$, if $\nu$ gives positive probability to all games in $\G$, and
$\K(\<\Gamma',I\>^*)=\{\<\Gamma',h\>:h\in I\}$ for every information
set $\<\Gamma',I\>^*$ of $\Gamma^{\nu}$, then $(\vec{\sigma}',\mu')$
is a generalized sequential equilibrium of $\Gamma^*$ iff
$(\vec{\sigma},\mu)$ is a conditional sequential equilibrium of
$\Gamma^{\nu}$ with respect to $\K$, where
$\sigma_{i,\Gamma'}(\<\Gamma^h\>\cdot
h')=\sigma'_{i,\Gamma'}(\Gamma^h,h')$ and $\mu'_{\Gamma',
I}=\mu_{\<\Gamma',I\>^*}$. \eothm

\prf Let $\Pr^{\nu}_{\vec{\sigma}}$ be the probability distribution
over the histories in $\Gamma^{\nu}$ induced by the strategy profile
$\vec{\sigma}$ and $f_c^{\nu}$. For a history $h$ of the game,
define $\Pr^{\nu}_{\vec{\sigma}}(\cdot\mid h)$ to be the conditional
probability distribution induced by $\vec{\sigma}$ and $f_c^{\nu}$
over the possible histories of the game given that the current
history is $h$. Similarly, let $\Pr^{h}_{\vec{\sigma}'}$ be the
probability distribution over the histories in $\Gamma^h\in \G$
induced by the generalized strategy profile $\vec{\sigma}'$ and
$f_c^h$. Note that if $\Pr^{h}_{\vec{\sigma}'}(h')>0$, then $h'\in
\lfloor H^h \rfloor$. Thus, $\<\Gamma^h\>\cdot h' \in H^{\nu}$.

For all strategy profiles $\sigma$ and generalized strategy profiles
$\sigma'$, if
$\sigma'_{i,\Gamma'}(\Gamma^h,h')=\sigma_{i,\Gamma'}(\<\Gamma^h\>\cdot
h')$, then it is easy to see that for all $h'\in H^h$ such that
$\Pr^{h}_{\vec{\sigma}'}(h')>0$, we have that
$\Pr^{\nu}_{\vec{\sigma}}(\<\Gamma^h\>\cdot
h')=\nu(\Gamma^h)\Pr^{h}_{\vec{\sigma}'}(h')$. And since $\nu$ is a
probability measure such that $\nu(\Gamma^h)>0$ for all $\Gamma^h\in
\G$, we have that $\Pr^{\nu}_{\vec{\sigma}}(\<\Gamma^h\>\cdot h')>0$
iff $\Pr^{h}_{\vec{\sigma}'}(h')>0$. It is also easy to see that for
all $h'\ne \< \, \>$ and all $h''\in H^h$ such that
$\Pr^{h}_{\vec{\sigma}'}(h''\mid h')>0$,
$\Pr^{\nu}_{\vec{\sigma}}(\<\Gamma^h\>\cdot h''\mid
h')=\Pr^{h}_{\vec{\sigma}'}(h''\mid h')$.

Suppose that $(\vec{\sigma},\mu)$ is a conditional sequential
equilibrium of $\Gamma^{\nu}$ with respect to $\K$. We first prove
that $(\vec{\sigma}',\mu')$ satisfies generalized sequential
rationality. Suppose, by way of contradiction, that it does not. Thus, there
exists a player $i$, a generalized $i$-information set
$(\Gamma^+,I)^*$, and a local strategy $s'$ for player $i$ in
$\Gamma^+$ such that
$$
\sum_{h\in I}\sum_{z\in
Z^+}\mu_{\Gamma^+,I}'(h)\Pr^+_{\vec{\sigma}'}(z\mid h)u_i^+(z)
< \sum_{h\in I}\sum_{z\in
Z^+}\mu_{\Gamma^+,I}'(h)\Pr^+_{(\vec{\sigma}'_{-(i,\Gamma')},s')}(z\mid
h)u_i^+(z).
$$

Define $s$ to be a strategy for player $(i,\Gamma^+)$ in
$\Gamma^{\nu}$ such that $s(\<\Gamma^h\>\cdot h')=s'(\Gamma^h,h')$.
Using the observation in the previous paragraph and the fact that
$\mu'_{\Gamma^+, I}=\mu_{\<\Gamma^+,I\>^*}$ and
$\K(\<\Gamma^+,I\>^*)=\{\<\Gamma^+,h\>:h\in I\}$, it follows that
\begin{eqnarray}\label{chap2:eq2}
& & \sum_{\<\Gamma^+,h\>\in \K(\<\Gamma^+,I\>^*)}\sum_{z\in \lfloor
Z^+\rfloor}\mu_{\<\Gamma^+,I\>^*}(h)\Pr^{\nu}_{\vec{\sigma}}(\<\Gamma^+\>\cdot
z\mid h)u_i^+(z) \nonumber \\
&  < &\sum_{\<\Gamma^+,h\>\in \K(\<\Gamma^+,I\>^*)}\sum_{z\in
\lfloor
Z^+\rfloor}\mu_{\<\Gamma^+,I\>^*}(h)\Pr^{\nu}_{(\vec{\sigma}_{-(i,\Gamma')},s)}(\<\Gamma^+\>\cdot
z\mid h)u_i^+(z). \nonumber \\
\end{eqnarray}

\noindent
By definition of $u_{i,\Gamma^+}^{\nu}$,
\commentout{
(\ref{chap2:eq2}) holds iff
\begin{eqnarray}\label{chap2:eq3}
& & \sum_{\<\Gamma^+,h\>\in \K(\<\Gamma^+,I\>^*)}\sum_{z^{\nu}\in
Z^{\nu}}\mu_{\<\Gamma^+,I\>^*}(h)\Pr^{\nu}_{\vec{\sigma}}(\<\Gamma^+\>\cdot
z\mid h)u_{i,\Gamma^+}^{\nu} \nonumber \\
& < &\sum_{\<\Gamma^+,h\>\in \K(\<\Gamma^+,I\>^*)}\sum_{z^{\nu}\in
Z^{\nu}}\mu_{\<\Gamma^+,I\>^*}(h)\Pr^{\nu}_{(\vec{\sigma}_{-(i,\Gamma')},s)}(\<\Gamma^+\>\cdot
z\mid h)u_{i,\Gamma^+}^{\nu}.\nonumber\end{eqnarray}
}
$u_i^+(z) = u_{i,\Gamma^+}^{\nu}(\<\Gamma^+\>,z)$.  Replacing
$u_i^+(z)$ by $u_{i,\Gamma^+}^{\nu}(\<\Gamma^+\>,z)$ in~(\ref{chap2:eq2}), it
follows that
$(\vec{\sigma},\mu)$ does not satisfy sequential
rationality in $\Gamma^\nu$, a contradiction. So,
$(\vec{\sigma}',\mu')$ satisfies generalized sequential rationality.
It remains to show that $\mu'$ is consistent
with $\vec{\sigma}'$.

Suppose that, for every generalized information set $(\Gamma^+,I)^*$,
$\sum_{h\in I}\Pr^+_{\vec{\sigma}'}(h)>0$. By definition of $\K$ and
the fact that for all $h'\in H^{\nu}$,
$\Pr^{\nu}_{\vec{\sigma}}(\<\Gamma^+\>\cdot h')>0$ iff
$\Pr^+_{\vec{\sigma}'}(h')>0$, we have that for every information set
$\<\Gamma^+,I\>^*$ of $\Gamma^{\nu}$,
$$\sum_{\<\Gamma^+,h\>\in
\K(\<\Gamma^+,I\>^*)}\Pr^{\nu}_{\vec{\sigma}}(\<\Gamma^+\>\cdot
h)>0.$$ Thus, by consistency of $\mu$, $\vec{\sigma}$, and $\K$, it
follows that for every information set $\<\Gamma^+,I\>^*$ of
$\Gamma^{\nu}$ and every $h\in \K(\<\Gamma^+,I\>^*)$, we have
$$\mu_{\<\Gamma^+,I\>^*}(h)=\frac{\Pr^{\nu}_{\vec{\sigma}}(\<\Gamma^+\>\cdot
h)}{\sum_{h'\in
\K(\<\Gamma^+,I\>^*)}\Pr^{\nu}_{\vec{\sigma}}(\<\Gamma^+\>\cdot
h')}.$$
Since $\mu'_{\Gamma^+, I}=\mu_{\<\Gamma^+,I\>^*}$,
$\K(\<\Gamma',I\>^*)=\{\<\Gamma',h\>:h\in I\}$, and for all $h'\in
H^h$ such that $\Pr^{h}_{\vec{\sigma}'}(h')>0$, we have that
$\Pr^{\nu}_{\vec{\sigma}}(\<\Gamma^h\>\cdot
h')=\nu(\Gamma^h)\Pr^{h}_{\vec{\sigma}'}(h')$, it is easy to see that
for every generalized information set $(\Gamma^+,I)^*$ and every
$h\in I$,
$$\mu'_{\Gamma^+,I}(h)=\frac{\Pr^+_{\vec{\sigma}'}(h)}{\sum_{h'\in
I}\Pr^+_{\vec{\sigma}'}(h')}.$$
Thus, $\mu'$ is consistent with $\vec{\sigma}'$.

Finally, suppose that there exists a generalized information set
$(\Gamma^+,I)^*$ such that $\sum_{h\in I}\Pr^+_{\vec{\sigma}'}(h)=0$.
By definition of $\K$ and the fact that for all $h'\in H^{\nu}$,
$\Pr^{\nu}_{\vec{\sigma}}(\<\Gamma^+\>\cdot h')>0$ iff
$\Pr^+_{\vec{\sigma}'}(h')>0$, we have that $\sum_{\<\Gamma^+,h\>\in
\K(\<\Gamma^+,I\>^*)}\Pr^{\nu}_{\vec{\sigma}}(\<\Gamma^+\>\cdot
h)=0$. Thus, by the consistency of $\mu$, $\vec{\sigma}$, and $\K$,
there exists a sequence of $\K$-assessments
$(\vec{\sigma}^n,\mu^n)$
such that $\vec{\sigma}^n$ consists of completely mixed strategies,
$\mu^n$ is consistent with $\vec{\sigma}^n$ and $\K$,
and $(\vec{\sigma}^n,\mu^n)$ converges pointwise to
$(\vec{\sigma},\mu)$.

Define a sequence of $\K$-assessments $(\vec{\tau}^{n},\nu^{n})$
such that $\nu^{n}_{\Gamma', I}=\mu^n_{\<\Gamma',I\>^*}$ and
$\sigma^n_{j,\Gamma'}(\<\Gamma^h\>\cdot h')=\tau^{n}
_{j,\Gamma'}(\Gamma^h,h')$ for all $n$. Since $\vec{\sigma}^n$
is completely mixed, so is $\vec{\tau}^n$; it also
follows from the earlier argument
that $\nu^{n}$ is consistent with $\vec{\tau}^{n}$ for all $n$.
Since $(\vec{\sigma}^n,\mu^n)$ converges
pointwise to $(\vec{\sigma},\mu)$, it is easy to see that
$(\vec{\tau}^{n},\nu^{n})$ converges pointwise to
$(\vec{\sigma}',\mu')$. Thus, $\mu'$ is consistent with
$\vec{\sigma}'$, and $(\vec{\sigma}',\mu')$ is a generalized
sequential equilibrium of $\Gamma^*$, as desired. The proof of the
converse is similar; we leave details to the reader. \eprf

\othm{chap2:thm:conn} For every extensive game $\Gamma$ with
countably many players where each player has finitely many pure
strategies and for every possibility system $\K$,
if $(\vec{\sigma},\mu)$ is a conditional sequential equilibrium of
$\Gamma$ with respect to $\K$, then there exists a belief system
$\mu'$ such that $(\vec{\sigma}, \mu')$ is a sequential equilibrium
of $\Gamma$. \eothm

\prf Since $(\sigma,\mu)$ is a conditional sequential equilibrium of
$\Gamma$ with respect to $\K$, by the consistency of $\mu$, $\sigma$,
and $\K$, there exists a sequence of $\K$-assessments
$(\hat{\sigma}^k,\hat{\mu}^k)$ such that $\hat{\sigma}^k$
is completely mixed,
$\hat{\mu}^k$ is consistent with
$\hat{\sigma}^k$ and $\K$, and $(\hat{\sigma}^k,\hat{\mu}^k)$
converges pointwise to $(\sigma,\mu)$. Let $\hat{\nu}^{k}$ be
the
belief system consistent with $\hat{\sigma}^k$. Using the same
techniques as in the proof of Lemma~\ref{chap2:lem:perfseq}, we
can construct a subsequence of $(\hat{\sigma}^k,\hat{\nu}^{k})$
that converges pointwise to $(\sigma,\mu')$. Thus, $\mu'$ is
consistent with $\sigma$. It remains to show that $(\sigma,\mu')$
satisfies sequential rationality.

Since, by definition of $\K$, for every $i$-information set $I$ of
$\Gamma$, player $i$ has the same utility for every run extending a
history in $I-\K(I)$, it is not hard to show that
$$\EU_i((\sigma,\mu')\mid I)=C + \mu'(K(I))\EU_i((\sigma,\mu)\mid
I)\mbox{,}$$ where $C$ and $\mu'(K(I))$ are independent of
$\sigma_i(I)$. Since, by sequential rationality, $\sigma_i(I)$
is a best response given $\mu$, it is also a best response given
$\mu'$. It follows that $(\sigma,\mu')$ is a sequential equilibrium
of $\Gamma$, as desired. \eprf

\subsection{PROOF OF THEOREM~\ref{chap2:rat_gNash}}
\othm{chap2:rat_gNash} If $\Gamma$ is a standard normal-form
game and $\vec{s}$ is a (pure) strategy profile such that for all
$i\in N$, $s_i$ is a correlated rationalizable strategy of player
$i$ in $\Gamma$, then
\begin{itemize}
\item[(i)] there is a (pure) generalized Nash equilibrium $\vec{s}^{\,*}$ of
$\Gamma^*(\vec{s})$ such that for every player $i$,
$\underline{s}^*_{i,\Gamma^{s_i}} =s_i$;

\item[(ii)] for every (pure) generalized Nash equilibrium $\vec{s}^{\,*}$ of
$\Gamma^*(\vec{s})$, for every local strategy $s^*_{i,\Gamma'}$ for
every player $i$ in $\vec{s}^{\,*}$,
the strategy $\underline{s}^*_{i,\Gamma'}$ is correlated
rationalizable for player $i$ in $\Gamma$.
\end{itemize}
\eothm

\prf Let $\Gamma^*(\vec{s})=(\G,\Gamma^m,\F)$ be as defined in
Section~\ref{chap2:sec:grat}. For part (i), consider the generalized
strategy profile $\vec{s}^{\,*}$ where, for every player $i$ and
every $\Gamma^{s'_i}\in \G_i$,
$i$ makes the same move according to both $s^*_{i,\Gamma^{s'_i}}$ and
$s'_i$.
Note that, by definition of $\Gamma^*(\vec{s})$, for all $h\in H_i^m$ we
have that $\F(\Gamma^m,h)=(\Gamma^{s_i},\cdot)$.
Thus, by definition of $\vec{s}^{\,*}$, for every player $i$,
$\underline{s}^*_{i,\Gamma^{s_i}} =s_i$.  It is easy to check
using the definition of $\Gamma^*(\vec{s})$, that $\vec{s}^{\,*}$ is a
generalized Nash equilibrium,  and that, for all $i\in N$, $s_i$
is a rationalizable strategy for player $i$; we leave details to the
reader.

\commentout{ For part (ii), let $\vec{s}^{\,*}$ be an arbitrary pure
generalized Nash equilibrium of $\Gamma^*(\vec{s})$. Consider any
local strategy $s^*_{i,\Gamma^{s'_i}}$ for player $i$ in
$\vec{s}^{\,*}$. From the fact that $\vec{s}^{\,*}$ is a (pure)
generalized Nash equilibrium, we have that
$$s^*_{i,\Gamma^{s'_i}}\in argmax_{s''_i}\EU_{i,\Gamma^{s'_i}}((s''_i,\vec{s}^{\,*}_{-(i,\Gamma^{s'_i})}))$$
for any $s''_i$ a (pure) local strategy for player $i$ in
$\Gamma^{s'_i}$.

Let $\cS$ be the set of rationalizable strategies in $\Gamma$. By
definition of $\Gamma^*(\vec{s})$, $s'_i$ is a rationalizable
strategy for player $i$ in $\Gamma$.  Let $\pi^{s'_i}$ be a
probability distribution over $\cS_{-i}$ for which $s'_i$ is a best
response. Let $\vec{s}^{\,1}, \ldots, \vec{s}^{\,m}$ be the strategy
profiles in $\cS_{-i}$ that get positive probability according to
$\pi^{s'_i}$. By definition, $\Gamma^{s'_i}$ is the game where
nature initially makes one of $m$ moves, say $c_1, \ldots, c_m$ (one
corresponding to each strategy that gets positive probability
according to $\pi^{s'_i}$), where the probability of move $c_j$ is
$\pi^{s'_i}(\vec{s}^{\,j})$. After nature's choice a copy of
$\Gamma$ is played.

By definition, if players follow the generalized strategy profile
$\vec{s}^{\,*}$ in $\Gamma^*(\vec{s})$, then every player $k\ne i$
uses local strategy $s^*_{k,\Gamma^{s^j_k}}$ in the histories that
follow nature's move $c_j$ in $\Gamma^{s'_i}$. Thus,
$$\EU_{i,\Gamma^{s'_i}}((s''_i,\vec{s}^{\,*}_{-(i,\Gamma^{s'_i})}))=\sum_{j=1}^m
\pi^{s'_i}(\vec{s}^{\,j})\EU_i(s''_i,(s^*_{k,\Gamma^{s^j_k}}:k\in N-\{i\})).$$
}

For part (ii), let $\D_i=\{\underline{s}^*_{i,\Gamma'}:\Gamma'\in
\G_i\}$, i.e., $\D_i$ consists of the strategies in the underlying
game $\Gamma$ corresponding to some local strategy of player $i$ in
$\Gamma^*(\vec{s})$. We claim that $\D_i\subseteq B(\D_{-i})$. To
see this, let $s^*_{i,\Gamma'}$ be any local strategy for player $i$
in $\vec{s}^{\,*}$. Since $\vec{s}^{\,*}$ is a generalized Nash
equilibrium, $s^*_{i,\Gamma'}$ is a best response to the local
strategies used by other players in $\Gamma'$. Note that, by
definition of $\D_i$, for every other player $j\ne i$, there is a
strategy $\underline{s}_{j,\Gamma'}\in \D_j$ corresponding to the
local strategy $s_{j,\Gamma'}$ player $j$ follows in game $\Gamma'$.
Since, by definition of $\Gamma^*(\vec{s})$, in game $\Gamma'$
nature makes an initial choice and then a copy of $\Gamma$ is
played, and all players but $i$ know the move made by nature, this
initial move by nature can be seen as a distribution over the local
strategies used by the other players in the different copies of
$\Gamma$ contained in $\Gamma'$. Thus, it is easy to see that the
strategy $\underline{s}_{i,\Gamma'}$ corresponding to
$s^*_{i,\Gamma'}$ is in $B(\D_{-i})$. Finally, since
$s^*_{i,\Gamma'}$ is an arbitrary local strategy of player $i$ in
$\vec{s}^{\,*}$, it follows that
$\D_i\subseteq B(\D_{-i})$.  By the definition of correlated
rationalizable strategies, it follows that $\D_i\subseteq \cS_i$.
Thus, for player $i$ in $\Gamma(\vec{s})$ and every local strategy
$s^*_{i,\Gamma'}$ for $i$ in $\vec{s}^{\,*}$,
$\underline{s}^*_{i,\Gamma'}$ is correlated rationalizable for
player $i$ in $\Gamma$, as desired. \eprf

}

\end{document}

%% file: defn.tex
\newtheorem{THEOREM}{Theorem}[section]
\newenvironment{theorem}{\begin{THEOREM} \hspace{-.85em} {\bf :} }%
                        {\end{THEOREM}}
\newtheorem{LEMMA}[THEOREM]{Lemma}
\newenvironment{lemma}{\begin{LEMMA} \hspace{-.85em} {\bf :} }%
                      {\end{LEMMA}}
\newtheorem{COROLLARY}[THEOREM]{Corollary}
\newenvironment{corollary}{\begin{COROLLARY} \hspace{-.85em} {\bf :} }%
                          {\end{COROLLARY}}
\newtheorem{PROPOSITION}[THEOREM]{Proposition}
\newenvironment{proposition}{\begin{PROPOSITION} \hspace{-.85em} {\bf :} }%
                            {\end{PROPOSITION}}
\newtheorem{DEFINITION}[THEOREM]{Definition}
\newenvironment{definition}{\begin{DEFINITION} \hspace{-.85em} {\bf :} \rm}%
                            {\end{DEFINITION}}
\newtheorem{CLAIM}[THEOREM]{Claim}
\newenvironment{claim}{\begin{CLAIM} \hspace{-.85em} {\bf :} \rm}%
                            {\end{CLAIM}}
\newtheorem{EXAMPLE}[THEOREM]{Example}
\newenvironment{example}{\begin{EXAMPLE} \hspace{-.85em} {\bf :} \rm}%
                            {\end{EXAMPLE}}
\newtheorem{REMARK}[THEOREM]{Remark}
\newenvironment{remark}{\begin{REMARK} \hspace{-.85em} {\bf :} \rm}%
                            {\end{REMARK}}

\newcommand{\thm}{\begin{theorem}}
\newcommand{\lem}{\begin{lemma}}
\newcommand{\pro}{\begin{proposition}}
\newcommand{\dfn}{\begin{definition}}
\newcommand{\rem}{\begin{remark}}
\newcommand{\xam}{\begin{example}}
\newcommand{\cor}{\begin{corollary}}
\newcommand{\prf}{\noindent{\bf Proof:} }
\newcommand{\ethm}{\end{theorem}}
\newcommand{\elem}{\end{lemma}}
\newcommand{\epro}{\end{proposition}}
\newcommand{\edfn}{\bbox\end{definition}}
\newcommand{\erem}{\bbox\end{remark}}
\newcommand{\exam}{\bbox\end{example}}
\newcommand{\ecor}{\end{corollary}}
\newcommand{\eprf}{\bbox\vspace{0.1in}}
\newcommand{\beqn}{\begin{equation}}
\newcommand{\eeqn}{\end{equation}}

\newcommand{\bbox}{\vrule height7pt width4pt depth1pt}

\newcommand{\clm}{\begin{claim}}
\newcommand{\eclm}{\end{claim}}

\newcommand{\bor}{\bigvee}

\newcommand{\union}{\cup}
\newcommand{\inter}{\cap}

\renewcommand{\phi}{\varphi}

\newcommand{\A}{{\cal A}}

\newcommand{\C}{{\cal C}}
\newcommand{\D}{{\cal D}}

\newcommand{\F}{{\cal F}}
\newcommand{\G}{{\cal G}}
\newcommand{\I}{{\cal I}}

\newcommand{\K}{{\cal K}}
\newcommand{\M}{{\cal M}}

\newcommand{\<}{\langle}
\renewcommand{\>}{\rangle}

\newcommand{\ol}{\setlength{\itemsep}{0pt}\begin{enumerate}}
\newcommand{\eol}{\end{enumerate}\setlength{\itemsep}{-\parsep}}
\newcommand{\ul}{\setlength{\itemsep}{0pt}\begin{itemize}}
\newcommand{\dl}{\setlength{\itemsep}{0pt}\begin{description}}
\newcommand{\edl}{\end{description}\setlength{\itemsep}{-\parsep}}
\newcommand{\eul}{\end{itemize}\setlength{\itemsep}{-\parsep}}

\newcommand{\cS}{{\cal S}}

\newcommand{\commentout}[1]{}

\newcommand{\bi}{\begin{itemize}}
\newcommand{\ei}{\end{itemize}}
\newcommand{\be}{\begin{enumerate}}
\newcommand{\ee}{\end{enumerate}}

%% file: bghkmac.tex
\newenvironment{oldthm}[1]{\par\noindent{\bf Theorem #1:} \em \noindent}{\par}
\newenvironment{oldlem}[1]{\par\noindent{\bf Lemma #1:} \em \noindent}{\par}
\newenvironment{oldcor}[1]{\par\noindent{\bf Corollary #1:} \em \noindent}{\par}
\newenvironment{oldpro}[1]{\par\noindent{\bf Proposition #1:} \em \noindent}{\par}
\newcommand{\othm}[1]{\begin{oldthm}{\ref{#1}}}
\newcommand{\eothm}{\end{oldthm} \medskip}
\newcommand{\olem}[1]{\begin{oldlem}{\ref{#1}}}
\newcommand{\eolem}{\end{oldlem} \medskip}
\newcommand{\ocor}[1]{\begin{oldcor}{\ref{#1}}}
\newcommand{\eocor}{\end{oldcor} \medskip}
\newcommand{\opro}[1]{\begin{oldpro}{\ref{#1}}}
\newcommand{\eopro}{\end{oldpro} \medskip}

\newcommand{\tends}{\rightarrow}
\newcommand{\tendsto}{\tends}

\newcommand{\bxor}[1]{\dot{\bor}}

